\documentclass[journal]{IEEEtran}
\usepackage{balance}
\usepackage{graphicx}
\usepackage{bm}
\usepackage{amsmath}
\usepackage{amsfonts}
\usepackage{amssymb}
\usepackage{color}
\usepackage{mathrsfs}
\usepackage{upgreek}
\usepackage[utf8]{inputenc}
\usepackage{tikz}
\usetikzlibrary{arrows, calc, decorations.markings, positioning}
\usepackage{cite}
\usepackage{booktabs}
\usepackage{multirow}
\usepackage{amsmath}
\usepackage{algorithm}
\usepackage{algorithmic}
\usepackage{makecell}
\usepackage[flushleft]{threeparttable}
\usepackage{tabularx}
\usepackage{threeparttable}
 \usepackage[caption=false,font=footnotesize]{subfig}
\usepackage{amssymb}
\usepackage{amsmath}
\usepackage{overpic}
\usepackage{algorithm}
\usepackage{algorithmic}
\usepackage{multirow}
\usepackage{array}
\usepackage{graphicx}
\usepackage{bm}
\usepackage{color}
\usepackage{tikz}
\usepackage{epstopdf}
\usepackage{stfloats}
\usepackage{cite}
\usepackage{tikz}
\usepackage{amsmath}
\usepackage{psfrag}
\usepackage{acronym}
\usepackage{enumerate}
\usetikzlibrary{arrows}
\usetikzlibrary{decorations}

\definecolor{BLUE}{rgb}{0,0,1}

\acrodef{aoa}[AOA]{angle-of-arrival}
\acrodef{bcrb}[BCRB]{Bayesian Cram\'{e}r-Rao bound}
\acrodef{bfim}[BFIM]{Bayesian Fisher information matrix}
\acrodef{bp}[BP]{belief propagation}
\acrodef{cdi}[CDI]{cooperative dilution intensity}
\acrodef{cir}[CIR]{channel impulse response}
\acrodef{cl}[CL]{cooperative localization}
\acrodef{cp}[CP]{cyclic prefix}
\acrodef{crb}[CRB]{Cram\'{e}r-Rao bound}
\acrodef{crlb}[CRLB]{Cram\'{e}r-Rao lower bound}
\acrodef{dft}[DFT]{discrete Fourier transform}
\acrodef{dof}[DoF]{degree of freedom}
\acrodef{dpeb}[DPEB]{directional position error bound}
\acrodef{fim}[FIM]{Fisher information matrix}
\acrodef{efim}[EFIM]{equivalent Fisher information matrix}
\acrodef{hogs}[HOGS]{hybrid orthogonal-Gaussian signalling}
\acrodef{hsug}[HSUG]{hybrid semi-unitary--Gaussian}
\acrodef{ici}[ICI]{information coupling intensity}
\acrodef{icrb}[ICRB]{inverse CRB}
\acrodef{iid}[i.i.d.]{independently and identically distributed}
\acrodef{im}[IM]{index modulation}
\acrodef{isac}[ISAC]{integrated sensing and communication}
\acrodef{los}[LoS]{line-of-sight}
\acrodef{mse}[MSE]{mean-squared error}
\acrodef{ofdm}[OFDM]{orthogonal frequency-division multiplexing}
\acrodef{pdf}[PDF]{probability density function}
\acrodef{peb}[PEB]{position error bound}
\acrodef{speb}[SPEB]{squared position error bound}
\acrodef{pll}[PLL]{phase-locked loop}
\acrodef{psk}[PSK]{phase shift keying}
\acrodef{p2p}[P2P]{point-to-point}
\acrodef{qam}[QAM]{quadrature amplitude modulation}
\acrodef{rbs}[RBS]{reference broadcast synchronization}
\acrodef{rhs}[RHS]{right hand side}
\acrodef{rii}[RII]{ranging information intensity}
\acrodef{rss}[RSS]{received signal strength}
\acrodef{rc}[RC]{ranging coefficient}
\acrodef{speb}[SPEB]{squared position error bound}
\acrodef{toa}[TOA]{time-of-arrival}
\acrodef{tdoa}[TDOA]{time-difference-of-arrival}
\acrodef{tpsn}[TPSN]{time synchronization protocol for sensor network}
\acrodef{vmp}[VMP]{variational message passing}
\acrodef{wsn}[WSN]{wireless sensor network}
\acrodef{efim}[EFIM]{equivalent Fisher information matrix}
\acrodef{dio}[DIO]{distance-information-only}
\acrodef{aio}[AIO]{angle-information-only}
\acrodef{saaf}[SAAF]{squared array aperture function}
\acrodef{snc}[S\&C]{sensing and communications}
\acrodef{uoa}[UOA]{uniformly oriented array}
\acrodef{rgg}[RGG]{random geometric graph}
\acrodef{snr}[SNR]{signal-to-noise ratio}
\acrodef{eoc}[EoC]{efficiency of cooperation}
\acrodef{npi}[NPI]{nominal position information}
\acrodef{gnss}[GNSS]{global navigation satellite system}
\acrodef{mimo}[MIMO]{multiple-input multiple-output}
\acrodef{mcs}[MCS]{minimally constrained system}
\acrodef{zzb}[ZZB]{Ziv-Zakai lower bound}
\acrodef{wwb}[WWB]{Weiss-Weinstein lower bound}
\acrodef{nlos}[NLOS]{non-light-of-sight}
\acrodef{mmse}[MMSE]{minimum mean squared error}
\acrodef{uav}[UAV]{unmanned aerial vehicle}
\acrodef{ppp}[PPP]{Poisson point process}
\acrodef{bpp}[BPP]{binomial point process}
\acrodef{cln}[CLN]{cooperative location-aware network}
\acrodef{pdr}[PDR]{pedestrian dead reckoning}
\acrodef{ml}[ML]{maximum likelihood}
\acrodef{map}[MAP]{maximum \textit{a posteriori}}
\acrodef{kkt}[KKT]{Karush-Kuhn-Tucker}
\acrodef{st}[ST]{subspace tradeoff}
\acrodef{drt}[DRT]{deterministic-random tradeoff}
\acrodef{ustm}[USTM]{unitary space-time modulation}
\acrodef{pri}[PRI]{pulse repetition interval}
\acrodef{prf}[PRF]{pulse repetition frequency}
\acrodef{acf}[ACF]{autocorrelation function}
\acrodef{isl}[ISL]{integrated sidelobe level}
\acrodef{islr}[ISLR]{integrated sidelobe level ratio}
\acrodef{rgi}[MFI]{mainlobe fluctuation intensity}
\acrodef{irgi}[AMFI]{averaged mainlobe fluctuation intensity}
\acrodef{cds}[CDS]{cyclic difference set}
\acrodef{psl}[PSL]{peak sidelobe level}
\acrodef{aesl}[AESL]{average expected sidelobe level}
\acrodef{pesl}[PESL]{peak expected sidelobe level}
\acrodef{masm}[MASM]{MASked Modulation}
\acrodef{si}[SI]{self-interference}
\acrodef{cpi}[CPI]{coherent processing interval}

\acrodefplural{dof}[DoFs]{degrees of freedom}
\acrodefplural{drt}[DRTs]{deterministic-random tradeoffs}
\acrodefplural{cds}[CDSs]{cyclic difference sets}

\usepackage{hyperref}					
\hypersetup{
	colorlinks=true,
	citecolor=blue,  
	linkcolor=blue,  
	urlcolor=blue    
}

\title{Dual Secure Sensing and Communications in MA-Enabled Networked ISAC Systems}
\author{Yan Yang, Jianwei Zhao, Zhendong Li,~\IEEEmembership{Member,~IEEE}, Wen Chen,~\IEEEmembership{Senior Member,~IEEE}, \\Weimin Jia, Feifei Gao,~\IEEEmembership{Fellow,~IEEE}, and Shi Jin,~\IEEEmembership{Fellow,~IEEE}
	\thanks{
		Y. Yang is with Rocket Force University of Engineering, Xi’an 710025, China, and also with the School of Information and Communication Engineering, Xi’an Jiaotong University, Xi’an 710049, China (e-mail: yangyanyy2022@163.com).
		J. Zhao and W. Jia are with Rocket Force University of Engineering, Xi’an 710025, China (e-mail: zhaojianweiep@163.com, jwm602@163.com). 
		Z. Li is with the School of Information and Communication Engineering, Xi’an Jiaotong University, Xi’an 710049, China (email: lizhendong@xjtu.edu.cn). 
		W. Chen is with the Department of Electronic Engineering, Shanghai Jiao Tong University, Shanghai 200240, China (e-mail: wenchen@sjtu.edu.cn). 
		F. Gao is with the Department of Automation, Tsinghua University, Beijing 100084, P. R. China (e-mail: feifeigao@ieee.org). 
		S. Jin is with the School of Information Science and Engineering, Southeast University, Nanjing 210096, China (email: jinshi@seu.edu.cn).
		(Corresponding authors: Jianwei Zhao; Zhendong Li)
	}
	\vspace{-7mm}
}

\begin{document}
\maketitle

\begin{abstract}	
Integrated sensing and communication (ISAC) is envisioned as a cornerstone technology for next-generation networks, yet it introduces significant security risks in both communication and sensing domains. To mitigate the coverage limitations inherent in conventional mono-static or bi-static setups, this paper proposes a networked ISAC framework empowered by movable antennas (MAs). By leveraging flexible antenna reconfiguration, MAs provide additional spatial degrees of freedom to significantly boost beamforming gain and interference suppression. Specifically, we investigate a novel dual-domain security paradigm where networked base stations collaboratively protect uplink (UL) and downlink (DL) transmissions from communication eavesdropping while simultaneously safeguarding target sensing data against sensing eavesdroppers (SEs). To this end, we formulate a joint optimization problem to maximize the sum of secrecy rates, subject to sensing requirements, SE eavesdropping probability constraint, and practical MA deployment constraints, by jointly optimizing receive combiners, transmit beamformers, UL power allocation, and MA positions. To address this non-convex problem with highly coupled variables, an efficient alternating optimization algorithm is proposed, integrating closed-form solutions for receive combiners, successive convex approximation for transmit beamforming and UL power, and a tailored sensing-aware multi-candidate particle swarm optimization for the high-dimensional MA positioning. Extensive numerical results validate the effectiveness of the proposed algorithms, demonstrating that the proposed MA-enabled networked cooperative ISAC design significantly outperforms conventional fixed-position antenna-based designs in terms of security performance, while achieving notable power- and antenna-saving gains.
\end{abstract}

\vspace{-2.5mm}
\vspace{-1mm}
\section{Introduction}

\IEEEPARstart{I}{ntegrated} sensing and communication (ISAC) has emerged as a pivotal enabler for sixth-generation (6G) wireless systems. By unifying radar sensing and communication, ISAC significantly enhances both hardware and spectral efficiency \cite{11250835}. However, the proliferation of emerging applications, such as smart cities, low-altitude economy, and intelligent transportation, imposes stringent demands for high-throughput data transmission and high-precision sensing that surpass the capabilities of conventional mono-static ISAC systems \cite{10616023}.

Against this backdrop, networked multi-static ISAC has garnered increasing interest. By interconnecting distributed base stations (BSs) via a central processor (CP), this architecture coordinates dispersed nodes for joint communication and sensing tasks \cite{11203895}. This coordination facilitates multi-view observation, extending spatial coverage and enhancing sensing accuracy to overcome the inherent limitations of mono-static configurations. Moreover, centralized joint signal processing can mitigate inter-cell interference and exploit macro-diversity gains. Accordingly, networked ISAC has attracted increasing research attention. For instance, \cite{10273396} outlined a multi-BS cooperative ISAC framework and elaborated on key enabling technologies.
In \cite{10516289}, a distributed massive multiple-input multiple-output (MIMO) ISAC system was investigated, introducing a communication-sensing region concept and proposing a dynamic resource allocation strategy. Furthermore, \cite{10892128} modeled distributed multi-node ISAC as a partial cooperative game to optimize grouping and beamforming, while \cite{10769538} derived analytical performance expressions for networked ISAC using stochastic geometry to reveal fundamental sensing-communication trade-offs.

While ISAC presents opportunities for enhanced service integration, it simultaneously introduces significant security challenges in both communication and sensing domains.
Due to the broadcast nature of wireless transmissions, ISAC systems are inherently vulnerable to potential communication security \cite{9199556,9755276,10536039,10746496} and sensing security threats \cite{10465106,8682745,10587082,10605793,11144762}.
In terms of communication security, sensing waveforms typically carry modulated information that could be intercepted by targets acting as potential eavesdroppers.
To address these issues, physical-layer security (PLS) techniques have been extensively explored.
For instance, the authors in \cite{9199556} incorporated artificial noise (AN) in a dual-functional system to degrade eavesdropping while ensuring the quality of service (QoS) for downlink (DL) legitimate users (LUs).
Further exploring the interplay among sensing, communication, and security, \cite{9755276} constructed an integrated network architecture that cohesively incorporates these functionalities.
In \cite{10536039}, secure transceiver design was investigated for full-duplex (FD) ISAC systems where sensing targets act as eavesdroppers.
Furthermore, \cite{10746496} examined a reconfigurable intelligent surface (RIS)-assisted FD networked ISAC system, balancing sensing and communication performance through the joint optimization of power allocation, beamforming, and RIS reflection coefficients.

Distinct from communication-only systems, ISAC introduces unique sensing security threats. Malicious sensing eavesdroppers (SEs) can passively intercept reflected signals to infer target behavior or environmental information without active transmission, thereby compromising sensing confidentiality \cite{10465106}, and may even deduce radar locations \cite{8682745}.
Sensing security refers to the protection of sensing-related information from such inference attacks, distinct from protecting data-bearing communication against communication eavesdroppers (CEs).
In sophisticated attack models, CEs and SEs may operate individually or collaboratively, resulting in dual-eavesdropping attacks that further aggravate security risks.
Several studies have preliminarily explored secure sensing and communication co-design. 
For instance, \cite{10587082} investigated an AN-aided beamforming design to mitigate information leakage to SEs while ensuring communication QoS.
Furthermore, \cite{10605793} optimized transmit beamforming for a networked secure ISAC system to maximize detection probability under dual eavesdropping threats, while \cite{11144762} addressed dual eavesdropping in a simultaneously transmitting and reflecting (STAR)-RIS-assisted ISAC network via joint beamforming and phase-shift optimization.
Despite these efforts, joint communication and sensing security in networked ISAC remains insufficiently explored.

The aforementioned studies rely on traditional fixed-position antennas (FPAs), which inherently limit the exploitation of channel spatial degrees of freedom (DoFs).
To overcome this limitation, movable antenna (MA) technology enables flexible antenna positioning within a continuous region, thereby providing additional spatial DoFs for channel reconfiguration \cite{10318061,10643473}. 
The work in \cite{10318061} established the multipath field-response channel model for MA-assisted communication systems, while subsequent studies demonstrated the benefits of MA reconfiguration for angle estimation \cite{10643473} and multi-user communications \cite{10741192,10354003,10694747}.
Recently, MA technology has also been explored in networked ISAC systems.
For example, \cite{10839251} enhanced both communication rate and sensing mutual information (MI) in a bi-static ISAC system by optimizing one-dimensional (1D) MA positions, while \cite{10944486} investigated two-dimensional (2D) MA positioning to maximize sensing signal-to-interference-plus-noise ratio (SINR). Furthermore, \cite{11204823} examined joint beamforming and MA positioning in a networked multi-static ISAC system, and \cite{2025arXiv250813839X} developed an MA-enhanced framework to mitigate hardware impairment distortions.
In the security domain, the application of MA technology shows promise by reconstructing channel responses to strengthen legitimate links while suppressing eavesdropping links, as demonstrated in various secure communication scenarios \cite{10416363,10684758,10818453}.
Specific to ISAC security, receiver-side \cite{11033708} and transmitter-side \cite{11020999} MA-assisted systems have been studied to improve secrecy performance. Despite these advances, the integration of MA reconfiguration into networked multi-static ISAC under joint communication and sensing security requirements has received limited attention.

However, existing studies remain largely separated across these research dimensions. Secure networked ISAC systems generally employ fixed antenna configurations, while MA-enabled ISAC mainly exploits spatial reconfigurability to enhance communication and/or sensing performance. Existing MA-assisted security designs, in turn, primarily focus on secure communications or secure ISAC with MA reconfiguration confined to either the transmitter or the receiver. Consequently, a networked multi-static architecture that simultaneously supports UL/DL communications, target sensing, coexisting communication and sensing eavesdroppers, and coordinated transmit/receive MA reconfiguration has yet to be systematically investigated.

The simultaneous incorporation of these capabilities gives rise to several distinct design challenges. First, transmit and receive MAs distributed across multiple BSs jointly reshape numerous communication, sensing, interference, and eavesdropping links. Hence, variations in MA positions can simultaneously affect multiple links and network functions, resulting in strong cross-node spatial coupling. Second, communication and sensing share common spatial DoFs and power resources and are jointly shaped by the same transceiver variables, while their operation is further constrained by communication- and sensing-security requirements. This creates strong cross-functional coupling among communication, sensing, and security objectives, such that improving one aspect may alter the channel conditions or resource allocation available to the others. Third, coordinated transmit/receive MA reconfiguration leads to a high-dimensional position-dependent design problem, in which the spatial configurations of distributed BSs are tightly coupled with transceiver processing and power allocation under practical MA deployment constraints.

These characteristics are not fully addressed by existing approaches. Conventional FPA-based secure ISAC lacks the spatial reconfigurability required to exploit position-domain DoFs, whereas existing MA-enabled designs typically consider only a subset of the above system dimensions. 
Therefore, exploiting network-level cooperation together with continuous MA reconfigurability requires a unified design that coordinates communication, sensing, security, and spatial reconfiguration across distributed transceivers. Motivated by these considerations, this work develops an MA-enabled networked multi-static secure ISAC framework that jointly optimizes transceiver processing, power allocation, and transmit/receive MA positions to enhance communication and sensing security while satisfying the sensing performance requirements. The main contributions of this paper are summarized as follows:
\begin{itemize}
	\item  
	To the best of our knowledge, this is the first work to investigate an MA-enabled networked secure ISAC system with coexisting communication and sensing eavesdroppers. We develop a unified architecture in which distributed transmitting and receiving base stations (TBSs/RBSs) equipped with MAs cooperatively support multiple UL and DL LUs, perform target sensing, and mitigate eavesdropping from multiple CEs and SEs. In particular, the transmit MAs at multiple TBSs and the receive MAs at multiple RBSs are coordinately reconfigured to exploit network-wide spatial DoFs. We formulate a sum of secrecy rates (SSR) maximization problem by jointly optimizing receive combiners, beamformers, AN covariance, UL power allocation, and transmit/receive MA positions, subject to constraints on the maximum transmit power, sensing MI requirements, SE detection probability, and MA movement limits.

    \item 
    To tackle the resulting highly coupled non-convex optimization problem, an efficient alternating optimization (AO)-based algorithm is proposed. The original problem is decomposed into three tractable subproblems and solved iteratively. Specifically, closed-form expressions are derived for the sensing and UL receive combiners based on generalized Rayleigh quotients. The complex coupled variables of transmit beamforming, AN covariance, and UL power are jointly optimized via difference-of-convex (DC) reformulation and successive convex approximation (SCA). Furthermore, to address the high-dimensional MA-position subproblem, a sensing-aware multi-candidate particle swarm optimization (SA-MCPSO) algorithm is introduced to jointly optimize the transmit and receive MA positions under the practical MA deployment constraints.

   \item
    Extensive simulations validate that the proposed framework achieves superior secrecy performance compared with the benchmark schemes. The results demonstrate that optimized MA positioning can effectively reconfigure the spatial channel conditions to strengthen legitimate communication and sensing links while suppressing eavesdropping links, thereby improving the achievable SSR while satisfying the sensing requirements. Moreover, the proposed scheme exhibits notable power- and antenna-saving gains, achieving a given SSR with lower transmit power or fewer antennas than the benchmark schemes. These results confirm the effectiveness of exploiting MA-enabled spatial DoFs for improving both security performance and resource efficiency in networked ISAC systems.
\end{itemize}

The rest of this paper is organized as follows. In Section~\ref{2}, we introduce the system model and the associated problem formulation. Section~\ref{3} elaborates on the proposed AO algorithm for solving the optimization problem. Simulation results and comprehensive discussions are presented in Section~\ref{4}, followed by the conclusions in Section~\ref{5}.

\textit{Notations:} Boldface lowercase and uppercase letters denote vectors and matrices, respectively. $[\mathbf{a}]_i$ represents the $i$-th element of vector $\mathbf{a}$, while $\mathbf{A} \succeq 0$ indicates that $\mathbf{A}$ is positive semidefinite. The superscripts $(\cdot)^T$ and $(\cdot)^H$ denote the transpose and conjugate transpose, respectively. $|\cdot|$, $\|\cdot\|_2$, $\mathrm{Tr}(\cdot)$, and $\mathrm{rank}(\cdot)$ denote the absolute value, Euclidean norm, trace, and rank, respectively. $\odot$ and $\otimes$ represent the Hadamard and Kronecker products, respectively. The sets of $M \times N$ complex and real matrices are denoted by $\mathbb{C}^{M \times N}$ and $\mathbb{R}^{M \times N}$, respectively, and $\mathbf{I}_N$ denotes the identity matrix of order $N$. $\mathcal{CN}(\mathbf{0}, \mathbf{\Lambda})$ represents the circularly symmetric complex Gaussian (CSCG) distribution with zero mean and covariance matrix $\mathbf{\Lambda}$. $\sim$ and $\triangleq$ stand for ``distributed as'' and ``defined as'', respectively. $[x]^+ \triangleq \max\{0, x\}$ is omitted in subsequent derivations as it does not influence the optimization.

\vspace{-2.2mm}
\section{System Model and Problem Formulation}\label{2}

\begin{figure}[t!]
		\centering
		\includegraphics[width=70mm]{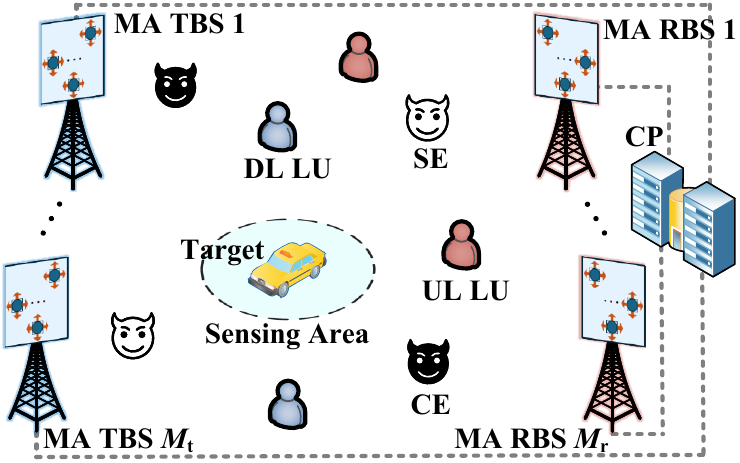}
		\vspace{-0.5em}
	\caption{Illustration of the MA-enabled networked secure ISAC system. }
	\label{fig:system_model_main}
	\vspace{-1em}
\end{figure}

As illustrated in Fig.~\ref{fig:system_model_main}, we consider an MA-enabled networked multi-static secure ISAC system. The system comprises $M_\mathrm{t}$ ISAC TBSs, $M_\mathrm{r}$ RBSs, $D$ single-antenna DL LUs, $U$ single-antenna UL LUs, and a point target. Additionally, the system operates in the presence of $E$ single-antenna CEs and $S$ SEs. The TBSs and RBSs are equipped with 2D planar arrays consisting of $N_\mathrm{t}=N_{\mathrm{t},x}\times N_{\mathrm{t},y}$ and $N_\mathrm{r}=N_{\mathrm{r},x}\times N_{\mathrm{r},y}$ MAs, respectively. Here, $N_{\mathrm{t},x}$ ($N_{\mathrm{r},x}$) and $N_{\mathrm{t},y}$ ($N_{\mathrm{r},y}$) denote the number of MAs along the horizontal and vertical axes, respectively. In contrast, each SE employs a uniform planar array (UPA) equipped with $N_\mathrm{S}$ FPAs.
In this ISAC system, the objective is to ensure secure communications for both $D$ DL LUs and $U$ UL LUs while simultaneously performing target sensing within a specific region of interest. In addition, it is assumed that all TBSs and RBSs are perfectly synchronized and connected to a CP via high-speed backhaul links to facilitate joint signal processing.

\vspace{-1.250em}
\subsection{Channel Model}
The position of the $n_\mathrm{t}$-th MA at TBS $m$ is denoted by $\mathbf{t}_{m,n_\mathrm{t}}^{\mathrm{t}}=[x_{m,n_\mathrm{t}}^{\mathrm{t}},y_{m,n_\mathrm{t}}^{\mathrm{t}}]^{T}\in\mathcal{C}_m^{\mathrm{t}}$, while that of the $n_\mathrm{r}$-th MA at RBS $p$ is given by $\mathbf{r}_{p,n_\mathrm{r}}^{\mathrm{r}}=[x_{p,n_\mathrm{r}}^{\mathrm{r}},y_{p,n_\mathrm{r}}^{\mathrm{r}}]^{T}\in\mathcal{C}_p^{\mathrm{r}}$. Here, $\mathcal{C}_m^{\mathrm{t}}$ and $\mathcal{C}_p^{\mathrm{r}}$ represent the designated local 2D continuous moving regions of size $A\times A$ for the transmit and receive domains, respectively. Given that the dimensions of the MA movement regions are significantly smaller than the signal propagation distances, the far-field channel model is adopted for all links between the BSs and LUs. Under this assumption, the specific positions of the MAs influence only the phase of the complex path coefficients, whereas the angles of departure (AoDs), angles of arrival (AoAs), and path amplitudes are assumed to remain constant relative to the reference point.

\subsubsection{Communication Channel Model}

Let $L_{m,d}^\mathrm{t}$, $L_{m,e}^\mathrm{t}$, and $L_{p,u}^\mathrm{r}$ denote the number of channel paths from TBS $m$ to DL LU $d$, from TBS $m$ to CE $e$, and from UL LU $u$ to RBS $p$, respectively. Focusing on the DL transmission, the elevation and azimuth AoDs of the $i$-th path from TBS $m$ to a receiver $\varrho$ are denoted by $\theta_{m,\varrho,i}^\mathrm{t}\in[-{\pi}/{2},{\pi}/{2}]$ and $\phi_{m,\varrho,i}^\mathrm{t}\in[-{\pi},{\pi}]$, respectively, where $\varrho\in\{d,e\}$ serves as a unified index for the DL LU $d$ and CE $e$. Consequently, the difference of the signal propagation distance for the $i$-th transmit path between the $n_\mathrm{t}$-th MA at position $\mathbf{t}_{m,n_\mathrm{t}}^\mathrm{t}$ and the reference point $\mathbf{t}_{m,0}^\mathrm{t}=[0,0]^T$ is expressed as
\begin{equation}\begin{aligned}&\rho_{m,\varrho,n_\mathrm{t},i}^\mathrm{t}(\mathbf{t}_{m,n_\mathrm{t}}^\mathrm{t})\\&=x_{m,n_\mathrm{t}}^\mathrm{t}\sin(\theta_{m,\varrho,i}^\mathrm{t})\cos(\phi_{m,\varrho,i}^\mathrm{t})+y_{m,n_\mathrm{t}}^\mathrm{t}\cos(\theta_{m,\varrho,i}^\mathrm{t}).\end{aligned}\end{equation}

Based on this, the field-response vector (FRV) capturing the channel response of the $n_\mathrm{t}$-th MA at TBS $m$ for the link to receiver $\varrho$ is given by
\begin{equation}\begin{aligned}\mathbf{h}_{m,\varrho,n_\mathrm{t}}^\mathrm{t}\!=\![e^{j\frac{2\pi}{\lambda}\rho_{m,\varrho,n_\mathrm{t},1}^\mathrm{t}},\cdots,e^{j\frac{2\pi}{\lambda}\rho_{m,\varrho,n_\mathrm{t},L_{m,\varrho}^\mathrm{t}}^\mathrm{t}}]^{T},\end{aligned}\end{equation}
where $\lambda$ denotes the carrier wavelength. 
By stacking the FRVs of all $N_\mathrm{t}$ MAs, the field-response matrix (FRM) for TBS $m$ is constructed as
\begin{equation}\mathbf{H}_{m,\varrho}^\mathrm{t}(\mathbf{t}_m^\mathrm{t})\!=\![\mathbf{h}_{m,\varrho,1}^\mathrm{t},\mathbf{h}_{m,\varrho,2}^\mathrm{t},\cdots,\mathbf{h}_{m,\varrho,N_\mathrm{t}}^\mathrm{t}]\in\mathbb{C}^{L_{m,\varrho}^\mathrm{t}\times N_\mathrm{t}},\end{equation}
where $\mathbf{t}_m^\mathrm{t}=[(\mathbf{t}_{m,1}^\mathrm{t})^T,\cdots,(\mathbf{t}_{m,N_\mathrm{t}}^\mathrm{t})^T]^T\in\mathbb{R}^{2N_\mathrm{t}\times1}$ collects the coordinates of all $N_\mathrm{t}$ transmit MAs at TBS $m$. Similarly, for the UL transmission, let $\theta_{p,u,i}^\mathrm{r}\in[-{\pi}/{2},{\pi}/{2}]$ and $\phi_{p,u,i}^\mathrm{r}\in[-{\pi},{\pi}]$ represent the elevation and azimuth AoAs of the $i$-th path from UL LU $u$ to RBS $p$. The corresponding difference of the signal propagation distance relative to the reference point for the $n_\mathrm{r}$-th MA at RBS $p$ is given by
\begin{equation}\begin{aligned}&\rho_{p,u,n_\mathrm{r},i}^\mathrm{r}(\mathbf{r}_{p,n_\mathrm{r}}^\mathrm{r})\\&=x_{p,n_\mathrm{r}}^\mathrm{r}\sin(\theta_{p,u,i}^\mathrm{r})\cos(\phi_{p,u,i}^\mathrm{r})+y_{p,n_\mathrm{r}}^\mathrm{r}\cos(\theta_{p,u,i}^\mathrm{r}).\end{aligned}\end{equation}

Then, the receive FRV of the $n_\mathrm{r}$-th MA at RBS $p$ for the channel from UL LU $u$ is given by
\begin{equation}\mathbf{h}_{p,u,n_\mathrm{r}}^\mathrm{r}=[e^{j\frac{2\pi}{\lambda}\rho_{p,u,n_\mathrm{r},1}^\mathrm{r}},\cdots,e^{j\frac{2\pi}{\lambda}\rho_{p,u,n_\mathrm{r},L_{p,u}^\mathrm{r}}^\mathrm{r}}]^{T}.\end{equation}

By aggregating the FRVs corresponding to all $N_\mathrm{r}$ receive MAs, the receive FRM for RBS $p$ is constructed as
\begin{equation}\mathbf{H}_{p,u}^\mathrm{r}(\mathbf{r}_p^\mathrm{r})=[\mathbf{h}_{p,u,1}^\mathrm{r},\mathbf{h}_{p,u,2}^\mathrm{r},\cdots,\mathbf{h}_{p,u,N_\mathrm{r}}^\mathrm{r}]\in\mathbb{C}^{L_{p,u}^\mathrm{r}\times N_\mathrm{r}},\end{equation}
where $\mathbf{r}_p^\mathrm{r}=[(\mathbf{r}_{p,1}^\mathrm{r})^T,\cdots,(\mathbf{r}_{p,N_\mathrm{r}}^\mathrm{r})^T]^T\in\mathbb{R}^{2N_\mathrm{r}\times1}$ collects the coordinates of all $N_\mathrm{r}$ MAs at RBS $p$.

Since the LUs and eavesdroppers are equipped with conventional FPAs, the MA-dependent FRMs characterize only the transmit side of the TBSs for DL channels and the receive side of the RBSs for UL channels. Then, we define path-response vectors (PRVs) $\mathbf{\Psi}_{m,d}\in \mathbb{C}^{L_{m,d}^{\mathrm{t}}\times1}$, $\mathbf{\Psi}_{m,e}\in \mathbb{C}^{L_{m,e}^{\mathrm{t}}\times1}$, and $\mathbf{\Psi}_{p,u}\in \mathbb{C}^{L_{p,u}^{\mathrm{r}}\times1}$, which account for the path gains and phases from TBS $m$ to DL LU $d$, from TBS $m$ to CE $e$, and from UL LU $u$ to RBS $p$, respectively. By combining the PRVs with spatial field responses, the effective channels between TBS $m$ and receiver $\varrho$, and between UL LU $u$ and RBS $p$, are expressed as
\begin{subequations} 
	\begin{align}
		&\mathbf{h}_{m,\varrho}\left(\mathbf{t}_m^\mathrm{t}\right)=\left(\mathbf{H}_{m,\varrho}^\mathrm{t}(\mathbf{t}_m^\mathrm{t})\right)^{H}\mathbf{\Psi}_{m,\varrho}\in\mathbb{C}^{N_\mathrm{t}\times 1},\\
		&\mathbf{h}_{p,u}\left(\mathbf{r}_p^\mathrm{r}\right)=\left(\mathbf{H}_{p,u}^\mathrm{r}(\mathbf{r}_p^\mathrm{r})\right)^{H}\mathbf{\Psi}_{p,u}\in\mathbb{C}^{N_\mathrm{r}\times 1 }.
	\end{align}
\end{subequations} 

In addition to the BS-LU links, we define the interference channels originating from the UL LUs. Specifically, let $h^{\mathrm{UD}}_{u,d}\in\mathbb{C}^{1\times1}$, $h^{\mathrm{UE}}_{u,e}\in\mathbb{C}^{1\times1}$, and $\mathbf{h}^{\mathrm{US}}_{u,s}\in\mathbb{C}^{N_\mathrm{S}\times1}$ denote the channel from UL LU $u$ to DL LU $d$, to CE $e$, and to SE $s$, respectively.

\subsubsection{Sensing Channel Model}

For the sensing channel, we adopt a single-path model, acknowledging that multi-hop reflected signals may suffer from severe path loss. Consequently, we assume that only the line-of-sight (LoS) path from the TBS to target/clutters and reflecting from target/clutters to the RBS can be successfully detected. Let the index $c \in \{0,1,\dots,K_\mathrm{C}\}$ represent the sensing components, where $c=0$ denotes the target of interest and $c \in \{1,\dots,K_\mathrm{C}\}$ represents $K_\mathrm{C}$ clutter components. 
For the $n_{\mathrm t}$-th MA of TBS $m$, the difference of the signal propagation distance for the $c$-th component relative to the reference point $\mathbf{t}_{m,0}^{\mathrm{t}}=[0,0]^T$ is expressed as
\begin{equation}\begin{aligned}&\rho_{m,c,n_\mathrm{t}}^\mathrm{t}(\mathbf{t}_{m,n_\mathrm{t}}^\mathrm{t})\\&=x_{m,n_\mathrm{t}}^\mathrm{t}\sin(\theta_{m,c}^\mathrm{t})\cos(\phi_{m,c}^\mathrm{t})+y_{m,n_\mathrm{t}}^\mathrm{t}\cos(\theta_{m,c}^\mathrm{t}),\end{aligned}\end{equation}
where $\theta_{m,c}^\mathrm{t}$ and $\phi_{m,c}^\mathrm{t}$ denote the elevation and azimuth AoDs from TBS $m$ toward the $c$-th sensing component, respectively. Accordingly, the transmit steering vector of TBS $m$ directed towards the $c$-th component is given by
\begin{equation}\mathbf{g}_{m,c}^\mathrm{t}=[e^{j\frac{2\pi}{\lambda}\rho_{m,c,1}^\mathrm{t}},\cdots,e^{j\frac{2\pi}{\lambda}\rho_{m,c,N_\mathrm{t}}^\mathrm{t}}]^{T}.\end{equation}

Similarly, the receive steering vector at RBS $p$ for the $c$-th component is defined as $\mathbf{g}_{p,c}^\mathrm{r}=[e^{j\frac{2\pi}{\lambda}\rho_{p,c,1}^\mathrm{r}},\cdots,e^{j\frac{2\pi}{\lambda}\rho_{p,c,N_\mathrm{r}}^\mathrm{r}}]^{T}$,
where the corresponding propagation difference is determined by $\rho_{p,c,n_\mathrm{r}}^\mathrm{r}(\mathbf{r}_{p,n_\mathrm{r}}^\mathrm{r})=x_{p,n_\mathrm{r}}^\mathrm{r}\sin(\theta_{p,c}^\mathrm{r})\cos(\phi_{p,c}^\mathrm{r})+y_{p,n_\mathrm{r}}^\mathrm{r}\cos(\theta_{p,c}^\mathrm{r})$, with $\theta_{p,c}^\mathrm{r}$ and $\phi_{p,c}^\mathrm{r}$ denoting the corresponding elevation and azimuth AoAs at RBS $p$. 
Based on these steering vectors, the effective propagation vectors associated with the \(c\)-th sensing component at TBS \(m\) and RBS \(p\) are modeled as \(\tilde{\mathbf g}_{m,c}^{\mathrm t}=\beta_{m,c}^{\mathrm t}\mathbf g_{m,c}^{\mathrm t}\) and \(\tilde{\mathbf g}_{p,c}^{\mathrm r}=\beta_{p,c}^{\mathrm r}\mathbf g_{p,c}^{\mathrm r}\), respectively, where \(\beta_{m,c}^{\mathrm t}\) and \(\beta_{p,c}^{\mathrm r}\) characterize the corresponding complex propagation coefficients accounting for attenuation and phase shifts, and are assumed to be accurately estimated.

\vspace{-1em}
\subsection{Transmit Signal Model}

Regarding the signal transmission, each TBS $m$ serves $D$ independent DL information streams while simultaneously ensuring security. Let $\mathbf{s}\triangleq[s_{1},\cdots,s_{D}]^T\in\mathbb{C}^{D\times1}$ denote the information symbols for $D$ DL LUs, satisfying $\mathbb{E}\{\mathbf{s}\mathbf{s}^{H}\}=\mathbf{I}_D$. To proactively degrade the interception capabilities of potential eavesdroppers, TBS $m$ additionally generates an AN vector $\mathbf{v}_m\in\mathbb{C}^{N_{\mathrm{t}}\times1}$. We assume the AN satisfies $\mathbf{v}_m\sim\mathcal{CN}\left(\mathbf{0},\mathbf{R}_m^\mathbf{v}\right)$ and is uncorrelated with the information signals, i.e., $\mathbb{E}\{\mathbf{s}\mathbf{v}_m^H\}=\mathbf{0}$. Consequently, the transmitted signal from TBS $m$ can be expressed as
\begin{equation}\mathbf{x}_m={\sum}_{d=1}^{D}\mathbf{w}_{m,d}s_d+\mathbf{v}_m\in\mathbb{C}^{N_{\mathrm{t}}\times1},\forall m,\end{equation}
where $\mathbf{w}_{m,d}\in\mathbb{C}^{N_\mathrm{t}\times1}$ is the transmit beamforming vector for the DL LU $d$. 
For global representation, we define $\mathbf{x}\triangleq[\mathbf{x}_1^{T},\cdots,\mathbf{x}_{M_\mathrm{t}}^{T}]^{T}\in\mathbb{C}^{M_\mathrm{t}N_\mathrm{t}\times1}$ as the aggregated signal vector across all $M_\mathrm{t}$ TBSs.

For UL communication, the signal transmitted by UL LU $u$ is given by
\begin{equation}x_u^\mathrm{U}=\sqrt{p_u}s_u^\mathrm{U},\forall u,\end{equation}
where $p_u$ denotes the transmit power and $s_u^\mathrm{U}$ represents the information symbol of UL LU $u$. The symbols $\{s_u^\mathrm{U}\}$ are mutually uncorrelated with zero mean and unit variance.

\vspace{-1em}
\subsection{Communication Model}
The signal received at DL LU  $d$ can be expressed as
\begin{align}
		&y^\mathrm{D}_{d}={\sum}_{m=1}^{M_\mathrm{t}}\mathbf{h}_{m,d}^{H}\mathbf{x}_{m}+{\sum}_{u=1}^{U}h^{\mathrm{UD}}_{u,d}x_{u}^\mathrm{U}+n^{\mathrm{D}}_{d}=\mathbf{h}_{d}^{H}\mathbf{w}_{d}s_{d}\nonumber\\
        &+{\sum}_{d'\neq d}^{D}\mathbf{h}_{d}^{H}\mathbf{w}_{d'}s_{d'}+\mathbf{h}_{d}^{H}\mathbf{v}+{\sum}_{u=1}^{U}h^{\mathrm{UD}}_{u,d}x_{u}^\mathrm{U} + n^\mathrm{D}_{d},
\end{align}
where $\mathbf{h}_{d}\triangleq[\mathbf{h}_{1,d}^{T},\mathbf{h}_{2,d}^{T},\cdots,\mathbf{h}_{M_\mathrm{t},d}^{T}]^{T}\in\mathbb{C}^{M_\mathrm{t}N_\mathrm{t}\times1}$ denotes the aggregated channel vector from all TBSs to DL LU $d$, $\mathbf{w}_d\triangleq[\mathbf{w}_{1,d}^T,\cdots,\mathbf{w}_{M_\mathrm{t},d}^T]^T\in\mathbb{C}^{M_\mathrm{t}N_\mathrm{t}\times1}$ denotes the aggregated beamforming vector, and $\mathbf{v}\triangleq[\mathbf{v}_{1}^{T},\cdots,\mathbf{v}_{M_\mathrm{t}}^{T}]^{T}\in\mathbb{C}^{M_\mathrm{t}N_\mathrm{t}\times 1}$ represents the global AN vector. $n^\mathrm{D}_{d}\sim\mathcal{CN}(0,\sigma_{\mathrm{D},d}^{2})$ models the additive white Gaussian noise (AWGN) at DL LU $d$. 

At the receiving side, the signal captured by RBS $p$ can be formulated as
\setcounter{equation}{12}
\begin{equation}
	\begin{aligned}\label{RBS_receive_w/o_u}
		\mathbf{y}_{p}^\mathrm{U}&={\sum}_{u=1}^{U}\mathbf{h}_{p,u}x_u^\mathrm{U}+{\sum}_{m=1}^{M_\mathrm{t}}\alpha_{0}\tilde{\mathbf{g}}^\mathrm{r}_{p,0}(\tilde{\mathbf{g}}^\mathrm{t}_{m,0})^H\mathbf{x}_m\\&+{\sum}_{c=1}^{K_\mathrm{C}}{\sum}_{m=1}^{M_\mathrm{t}}\alpha_{c}\tilde{\mathbf{g}}^\mathrm{r}_{p,c}(\tilde{\mathbf{g}}^\mathrm{t}_{m,c})^H\mathbf{x}_m+\mathbf{n}^\mathrm{r}_{p},
	\end{aligned}
\end{equation}
where $\alpha_c\sim\mathcal{CN}(0,\sigma_c^2)$ denotes the complex reflection coefficient associated with the radar cross section (RCS) of the $c$-th sensing component, which is assumed to remain unchanged during one coherent processing interval. $\mathbf{n}_{p}^\mathrm{r}\sim\mathcal{CN}(\mathbf{0},\sigma_\mathrm{r}^{2}\mathbf{I})$ denotes the AWGN at RBS $p$. 

To extract the desired UL information and sensing data, RBS $p$ applies a set of $U+1$ linear receive filters $\mathbf{u}_{p,i}\in\mathbb{C}^{N_\mathrm{r}\times1}$, where $i\in\{0,1,\dots,u,\dots,U\}$ and $\mathbf{u}_{p,0}$ is dedicated to sensing. The filtered output of stream $i$ at RBS $p$ is given by $y_{p,i}^\mathrm{U}=\mathbf{u}_{p,i}^H\mathbf{y}_{p}^\mathrm{U}$. After centralized processing at the CP, the aggregated signal from all $M_\mathrm{r}$ RBSs for stream $i$ is
\begin{equation}
	\begin{aligned}\label{RBS_receive_u}
		&y_{i}^\mathrm{U}={\sum}_{p=1}^{M_\mathrm{r}}y_{p,i}^\mathrm{U}={\sum}_{u=1}^{U}\mathbf{u}_{i}^{H}\mathbf{h}_{u}x_{u}^\mathrm{U}\\
		&+{\sum}_{c=0}^{K_\mathrm{C}}\alpha_{c}\mathbf{u}_{i}^{H}\mathbf{g}^\mathrm{r}_{c}(\mathbf{g}^\mathrm{t}_{c})^{H}({\sum}_{d=1}^{D}\mathbf{w}_{d}s_{d}+\mathbf{v})+\mathbf{u}_{i}^{H}\mathbf{n}^\mathrm{r},
	\end{aligned}
\end{equation}
where we define the aggregate vectors $\mathbf{u}_i\triangleq[\mathbf{u}_{1,i}^T,\cdots,\mathbf{u}_{M_\mathrm{r},i}^T]^T$, $\mathbf{h}_{u}\triangleq[\mathbf{h}_{1,u}^{T},\cdots,\mathbf{h}_{M_\mathrm{r},u}^{T}]^{T}$, and $\mathbf{n}^\mathrm{r}\triangleq[(\mathbf{n}^\mathrm{r}_{1})^{T},\cdots,(\mathbf{n}^\mathrm{r}_{M_\mathrm{r}})^{T}]^{T}$. The global effective sensing vectors are defined as $\mathbf{g}^\mathrm{t}_{c}\triangleq[(\tilde{\mathbf{g}}^\mathrm{t}_{1,c})^T,\cdots,(\tilde{\mathbf{g}}^\mathrm{t}_{M_\mathrm{t},c})^T]^T$ and $\mathbf{g}^\mathrm{r}_{c}\triangleq[(\tilde{\mathbf{g}}^\mathrm{r}_{1,c})^T,\cdots,(\tilde{\mathbf{g}}^\mathrm{r}_{M_\mathrm{r},c})^T]^T$. 

Based on these signal models, the achievable rates for DL LU $d$ and UL LU $u$ are given by ${R}^\mathrm{D}_{d}=\log_2(1+\text{SINR}^\mathrm{D}_{d})$ and ${R}^\mathrm{U}_{u}=\log_2(1+\text{SINR}^\mathrm{U}_{u})$, respectively. The corresponding SINR for DL LU $d$ is derived as
\begin{equation}
	\begin{aligned}
		&\text{SINR}^\mathrm{D}_{d}\\
		&=\!\frac{\mathrm{Tr}\!\left\{\mathbf{H}_{d}\mathbf{W}_{d}\right\}}{\sum_{d'\neq d}^{D}\!\mathrm{Tr}\!\left\{\mathbf{H}_{d}\mathbf{W}_{d'}\right\}\!+\!\mathrm{Tr}\!\left\{\mathbf{H}_{d}\mathbf{V}\right\}\!+\!\sum_{u=1}^{U}\!p_{u}|h^{\mathrm{UD}}_{u,d}|^{2}\!+\!\sigma_{\mathrm{D},d}^{2}},
	\end{aligned}
\end{equation}
where $\mathbf{W}_{d}=\mathbf{w}_{d}\mathbf{w}_{d}^H\in\mathbb{C}^{M_\mathrm{t}N_\mathrm{t}\times M_\mathrm{t}N_\mathrm{t}}$, $\mathbf{V}\triangleq\mathbb{E}[\mathbf{v}\mathbf{v}^H]\in\mathbb{C}^{M_\mathrm{t}N_\mathrm{t}\times M_\mathrm{t}N_\mathrm{t}}$, and $\mathbf{H}_{d}=\mathbf{h}_{d}\mathbf{h}_{d}^H\in\mathbb{C}^{M_\mathrm{t}N_\mathrm{t}\times M_\mathrm{t}N_\mathrm{t}}$. Similarly, the SINR for UL LU $u$ is expressed in \eqref{SINR_UL_LU}, where $\mathbf{G}_c=\sigma_c\mathbf{g}^\mathrm{r}_{c}(\mathbf{g}^\mathrm{t}_{c})^{H}$ denotes the effective sensing channel matrix associated with the \(c\)-th scattering component.
\begin{figure*}[hb] 
	\centering
	\vspace{-1.2em}
	\hrulefill
	\begin{align}\label{SINR_UL_LU}
		\begin{aligned}	    
			\text{SINR}^\mathrm{U}_{u}
			=\frac{p_{u}\mathbf{u}_{u}^{H}\mathbf{h}_{u}\mathbf{h}_{u}^{H}\mathbf{u}_{u}}{\mathbf{u}_{u}^{H}\left(\sum_{u'\neq u}^{U}p_{u'}\mathbf{h}_{u'}\mathbf{h}_{u'}^H+\sum_{c=0}^{K_\mathrm{C}}\mathbf{G}_{c}\left(\sum_{d=1}^{D}\mathbf{W}_{d}+\mathbf{V}\right)\mathbf{G}^{H}_{c}+\sigma_\mathrm{r}^{2}\mathbf{I}\right)\mathbf{u}_{u}},
		\end{aligned}
	\end{align}
\end{figure*}
\begin{figure*}[hb] 
	\centering
	\vspace{-1.2em}
	\hrulefill
	\begin{align}\label{SINR_t}
		\begin{aligned}	    
			\text{SINR}_\mathrm{t}=\frac{\mathbf{u}_{0}^{H}\mathbf{G}_0(\sum_{d=1}^D\mathbf{W}_d+\mathbf{V})\mathbf{G}_0^{H}\mathbf{u}_{0}}{\mathbf{u}_{0}^{H}\left(\sum_{c=1}^{K_\mathrm{C}}(\mathbf{G}_c(\sum_{d=1}^D\mathbf{W}_d+\mathbf{V})\mathbf{G}_c^{H})+\sum_{u=1}^{U}p_{u}\mathbf{h}_{u}\mathbf{h}_{u}^H+\sigma_\mathrm{r}^{2}\mathbf{I}\right)\mathbf{u}_{0}},
		\end{aligned}
	\end{align}
\end{figure*}

\vspace{-1.2em}
\subsection{Sensing Model}
In this subsection, we elaborate on the cooperative sensing model employed within the networked ISAC system.
The CP aggregates the processed signals from all $M_\mathrm{r}$ RBSs to facilitate joint target detection. 
Leveraging the centralized architecture, all TBSs and RBSs are connected to the CP through fronthaul links, enabling data-level fusion and joint signal processing. With the assistance of the CP, the transmitted waveforms of all TBSs are available at the RBSs. Moreover, the direct TBS-to-RBS links are assumed to remain quasi-static within one coherent processing interval. Therefore, the direct-link interference caused by TBS transmissions can be estimated and mitigated based on the prior information obtained through the fronthaul connections.
Thus, with the received signal at RBS $p$ defined in \eqref{RBS_receive_u}, the sensing SINR for target detection is derived as \eqref{SINR_t}, presented at the bottom of this page.

From an information-theoretic perspective, the sensing MI is adopted as the primary performance metric to quantify the amount of environmental information extractable from the received observations \cite{10839251}. Specifically, the sensing MI is defined as the mutual dependence between the aggregated received signal $y_{0}^\mathrm{U}$ and the target response matrix $\mathbf{G}_{0}$, conditioned on the known transmit waveform $\mathbf{x}$ and the applied receive beamformer $\mathbf{u}_{0}$. Accordingly, this conditional MI is expressed as
\setcounter{equation}{17}
\begin{equation}
	{R}^\mathrm{S}=I\left(y_{0}^\mathrm{U};\mathbf{G}_{0}\mid\mathbf{u}_{0},\mathbf{x}\right)=\log_2(1+\text{SINR}_\mathrm{t}),
\end{equation}
where $\mathbf{G}_0=\sigma_0\mathbf{g}^\mathrm{r}_{0}(\mathbf{g}^\mathrm{t}_{0})^{H}$ represents the effective channel matrix for the target\footnote{For analytical tractability, the main formulation assumes ideal TBS-to-RBS direct-link interference cancellation and does not explicitly account for hardware impairments. The impact of residual direct-link interference (RDI) is evaluated in Section~\ref{4}. In practice, power-amplifier nonlinearities, oscillator phase noise, and imperfect radio-frequency (RF) calibration may introduce waveform distortions and impair coherent target-echo processing, potentially reducing the achievable sensing SINR and MI relative to their nominal predictions. Comprehensive hardware-aware modeling is left for future work.}.


\vspace{-0.5em}
\subsection{Communication Eavesdropping Model}

The signal received by the CE $e$ is expressed as
\begin{align}\label{b}
	\begin{aligned}
		&y_{e}^{\mathrm{E}}={\sum}_{m=1}^{M_\mathrm{t}}{\sum}_{d=1}^{D}\mathbf{h}_{m,e}^{H}\mathbf{w}_{m,d}s_{d}+{\sum}_{m=1}^{M_\mathrm{t}}\mathbf{h}_{m,e}^{H}\mathbf{v}_m\\&+{\sum}_{u=1}^{U}h^{\mathrm{UE}}_{u,e}\sqrt{p_u}s_u^{\mathrm{U}}+n_{e}^{\mathrm{E}},
	\end{aligned}
\end{align}
where $n^\mathrm{E}_{e}\sim\mathcal{CN}(0,\sigma_{\mathrm{E},e}^{2})$ models the AWGN at CE $e$.
For a secure design, we adopt a worst-case assumption regarding the interception capabilities of the eavesdroppers. Specifically, the CEs are assumed to possess perfect channel state information (CSI) and operate cooperatively to intercept the confidential information \cite{10818453}. Furthermore, utilizing advanced successive interference cancellation (SIC) techniques, each CE is assumed to be capable of perfectly decoding and canceling the interference from other LUs before attempting to decode the desired UL or DL signal. Accordingly, the achievable rates for CEs eavesdropping on UL LU $u$ and DL LU $d$ are defined as ${R}_{u}^{\mathrm{EU}}=\log_{2}\left(1+\text{SINR}_{u}^{\mathrm{EU}}\right)$ and ${R}_{d}^{\mathrm{ED}}=\log_{2}\left(1+\text{SINR}_{d}^{\mathrm{ED}}\right)$, respectively. The corresponding $\text{SINR}_{u}^{\mathrm{EU}}$ and $\text{SINR}_{d}^{\mathrm{ED}}$ are derived in \eqref{SINR_UL_IE} and \eqref{SINR_DL_IE}, where $\mathbf{h}_{e}\triangleq[\mathbf{h}_{1,e}^{T},\cdots,\mathbf{h}_{M_\mathrm{t},e}^{T}]^{T}\in\mathbb{C}^{M_\mathrm{t}N_\mathrm{t}\times1}$ denotes the aggregated channel vector from all $M_\mathrm{t}$ TBSs to CE $e$, and $\mathbf{H}_{e}=\mathbf{h}_{e}\mathbf{h}_{e}^{H}\in\mathbb{C}^{M_\mathrm{t}N_\mathrm{t}\times M_\mathrm{t}N_\mathrm{t}}$ represents the corresponding channel matrix.
\begin{figure*}[hb] 
	\centering
	\vspace{-1.2em}
	\hrulefill 
	\begin{subequations} 
		\begin{align}
			&\begin{aligned}\label{SINR_UL_IE}
				\text{SINR}_{u}^{\mathrm{EU}}\!\!=\!\!\sum_{e=1}^{E}\!\frac{\left|h^{\mathrm{UE}}_{u,e}\right|^{2}p_{u}}
				{\left|\sum_{m=1}^{M_\mathrm{t}}\!\mathbf{h}_{m,e}^{H}\mathbf{v}_m\right|^{2}\!+\!\sigma_{\mathrm{E},e}^{2}}\!\!=\!\!\sum_{e=1}^{E}\!\frac{\left|h^{\mathrm{UE}}_{u,e}\right|^{2}p_{u}}
				{\left|\mathbf{h}_{e}^{H}\mathbf{v}\right|^{2}\!+\!\sigma_{\mathrm{E},e}^{2}}
				{\operatorname*{=}}
				\frac{\sum_{e=1}^{E}\!\!\left(\left|h^{\mathrm{UE}}_{u,e}\right|^{2}p_{u}\prod_{e'\neq e}^{E}\!\!\left(\mathrm{Tr}\left\{\mathbf{H}_{e'}\mathbf{V}\right\}\!+\!\sigma_{\mathrm{E},e'}^{2}\right)\right)}
				{\prod_{e=1}^{E}\!\!\left(\mathrm{Tr}\left\{\mathbf{H}_{e}\mathbf{V}\right\}\!+\!\sigma_{\mathrm{E},e}^{2}\right)},
			\end{aligned}\\
			&\begin{aligned}\label{SINR_DL_IE}
				\text{SINR}_{d}^{\mathrm{ED}}\!\!=\!\!\sum_{e=1}^{E}\!\frac{\left|\sum_{m=1}^{M_\mathrm{t}}\!\mathbf{h}_{m,e}^{H}\mathbf{w}_{m,d}\right|^{2}}
				{\left|\sum_{m=1}^{M_\mathrm{t}}\!\mathbf{h}_{m,e}^{H}\mathbf{v}_m\right|^{2}\!+\!\sigma_{\mathrm{E},e}^{2}}
				\!\!=\!\!\sum_{e=1}^{E}\!\frac{\left|\mathbf{h}_{e}^{H}\mathbf{w}_{d}\right|^{2}}
				{\left|\mathbf{h}_{e}^{H}\mathbf{v}\right|^{2}\!+\!\sigma_{\mathrm{E},e}^{2}}
				{\operatorname*{=}}
				\frac{\sum_{e=1}^{E}\!\!\left(\!\mathrm{Tr}\left\{\mathbf{H}_{e}\mathbf{W}_{d}\right\}\!\prod_{e'\neq e}^{E}\!\!\left(\mathrm{Tr}\left\{\mathbf{H}_{e'}\mathbf{V}\right\}\!+\!\sigma_{\mathrm{E},e'}^{2}\right)\!\right)}
				{\prod_{e=1}^{E}\!\left(\mathrm{Tr}\left\{\mathbf{H}_{e}\mathbf{V}\right\}\!+\!\sigma_{\mathrm{E},e}^{2}\right)},
			\end{aligned}
		\end{align}
	\end{subequations}
\end{figure*}

\vspace{-1em}
\subsection{Sensing Eavesdropping Model}

Unlike the authorized RBSs, SEs do not possess prior knowledge of the transmitted signal $\mathbf{x}$ required for coherent sensing. 
We assume that each SE can effectively suppress the direct TBS-to-SE component through long-term background calibration and spatial filtering prior to target detection. Since this direct component is independent of the target presence, its suppression provides the SE with a more favorable detection condition and allows the subsequent detection model to focus on the target- and clutter-reflected components that carry target-presence information.
Accordingly, each SE employs a passive energy detector to infer the presence of the target \cite{10605793}. SE $s$ is assumed to have knowledge of the transmit antenna configuration and employs a receive beamformer $(\mathbf{g}_{s,0}^\mathrm{S}(\theta_{s,0},\phi_{s,0}))^H$ to focus on the potential target direction defined by the elevation angle $\theta_{s,0}$ and azimuth angle $\phi_{s,0}$. The receive steering vector for SE $s$ is modeled as
\begin{equation}\mathbf{g}_{s,c}^\mathrm{S}(\theta_{s,c},\phi_{s,c})\!=\!\mathbf{g}_{s,P,c}^\mathrm{S}(\theta_{s,c},\phi_{s,c})\otimes\mathbf{g}_{s,Q,c}^\mathrm{S}(\theta_{s,c},\phi_{s,c})\!\in\!\mathbb{C}^{N_\mathrm{S}\times1},\end{equation}
where $\mathbf{g}^\mathrm{S}_{s,Q}(\theta,\phi)=[1,\ldots,e^{j\frac{2\pi}{\lambda}d_Q(N_Q-1)\sin(\theta)\sin(\phi)}]^T$ and $\mathbf{g}^\mathrm{S}_{s,P}(\theta,\phi)=[1,\ldots,e^{j\frac{2\pi}{\lambda}d_P(N_P-1)\sin(\theta)\cos(\phi)}]^T$ denote the array response vectors for the horizontal and vertical dimensions of the SE's FPA, respectively. Let $\alpha_{s,m,c}\sim\mathcal{CN}(0,\sigma_{s,m,c}^2)$ represent the complex scattering coefficient characterizing the path from TBS $m$ to SE $s$ via the $c$-th scatterer. After receive beamforming, the signal observed by SE $s$ is given by
\begin{equation}
	\begin{aligned}
		&y_{s}\!=\!(\mathbf{g}_{s,0}^\mathrm{S})^H\bigg({\sum}_{m=1}^{M_\mathrm{t}}\alpha_{s,m,0}\tilde{\mathbf{g}}_{s,0}^\mathrm{S}(\tilde{\mathbf{g}}^\mathrm{t}_{m,0})^H\mathbf{x}_m\\&+\!{\sum}_{c=1}^{K_\mathrm{C}}{\sum}_{m=1}^{M_\mathrm{t}}\alpha_{s,m,c}\tilde{\mathbf{g}}_{s,c}^\mathrm{S}(\tilde{\mathbf{g}}^\mathrm{t}_{m,c})^H\mathbf{x}_m\!+\!{\sum}_{u=1}^{U}\mathbf{h}^{\mathrm{US}}_{u,s}x_u^\mathrm{U}\!+\!\mathbf{n}^\mathrm{S}_{s}\bigg),
	\end{aligned}
\end{equation}
where $\mathbf{n}^\mathrm{S}_{s}\sim\mathcal{CN}(\mathbf{0},\sigma_{\mathrm{S},s}^{2}\mathbf{I})$ denotes the additive noise. $\tilde{\mathbf{g}}^\mathrm{S}_{s,c}=\beta_{s,c}^\mathrm{S}\mathbf{g}_{s,c}^\mathrm{S}$ denotes the effective propagation vector from the $c$-th sensing component to SE $s$, where each $\beta_{s,c}^\mathrm{S}$ is a complex fading coefficient and is assumed to be known.
For convenience, the received signal can be rewritten as
\begin{equation}	
\begin{aligned}
		y_{s}=\mathbf{g}_s^H\mathbf{x}+\mathbf{u}_s^H\mathbf{x}+(\mathbf{g}^\mathrm{S}_{s,0})^H\sum_{u=1}^{U}\mathbf{h}^{\mathrm{US}}_{u,s}x_u^\mathrm{U}+(\mathbf{g}^\mathrm{S}_{s,0})^H\mathbf{n}^\mathrm{S}_{s},	
\end{aligned}
\end{equation}
where $\mathbf{g}_s\!\!=\!\![N_\mathrm{S}\alpha_{s,1,0}\beta_{s,0}^\mathrm{S}(\tilde{\mathbf{g}}^\mathrm{t}_{1,0})^H,\!\ldots\!,N_\mathrm{S}\alpha_{s,M_\mathrm{t},0}\beta_{s,0}^\mathrm{S}(\tilde{\mathbf{g}}^\mathrm{t}_{M_\mathrm{t},0})^H]^H$ represents the aggregated equivalent effective response vector for the target. $\mathbf{u}_s=\left[(\mathbf{g}^\mathrm{S}_{s,0})^H\mathbf{U}_{1,s},\ldots,(\mathbf{g}^\mathrm{S}_{s,0})^H\mathbf{U}_{M_\mathrm{t},s}\right]^H$ denotes the aggregated equivalent clutter response vector, where $\mathbf{U}_{m,s}=\sum_{c=1}^{K_\mathrm{C}}\alpha_{s,m,c}\tilde{\mathbf{g}}^\mathrm{S}_{s,c}(\tilde{\mathbf{g}}^\mathrm{t}_{m,c})^H$.  The detection problem at SE $s$ is formulated as a binary hypothesis test (i.e. null hypothesis $\mathcal{H}_{0}$, target absent, or alternative hypothesis $\mathcal{H}_{1}$, target present), which can be given as \cite{11203895}
\begin{equation}
	\begin{cases}
		\mathcal{H}_0:y_{s}=\mathbf{u}_s^H\mathbf{x}+{\sum}_{u=1}^{U}G_{u,s}x_u^\mathrm{U}+\hat{n}^\mathrm{S}_{s},\\
		\mathcal{H}_1:y_{s}=\mathbf{g}_s^H\mathbf{x}+\mathbf{u}_s^H\mathbf{x}+{\sum}_{u=1}^{U}G_{u,s}x_u^\mathrm{U}+\hat{n}^\mathrm{S}_{s},
		\end{cases}
\end{equation}
where $G_{u,s}=(\mathbf{g}^\mathrm{S}_{s,0})^H\mathbf{h}^{\mathrm{US}}_{u,s}$ represents the projected UL interference, and $\hat{n}^\mathrm{S}_{s}=(\mathbf{g}^\mathrm{S}_{s,0})^H\mathbf{n}^\mathrm{S}_{s}\sim\mathcal{CN}(0,N_\mathrm{S}\sigma_{\mathrm{S},s}^{2})$ is the post-processing noise.
Since $\mathbf{x}$ is unknown, the SE employs an energy detector $|y_{s}|^2$. The likelihood functions of $y_{s}$ under the two hypotheses follow complex Gaussian distributions:
\begin{equation}
	\begin{cases}
		p_0(y_{s})=\frac{1}{\pi\Omega_{0,s}}\exp(-\frac{|y_{s}|^2}{\Omega_{0,s}}),\\
		p_1(y_{s})=\frac{1}{\pi\Omega_{1,s}}\exp(-\frac{|y_{s}|^2}{\Omega_{1,s}}),
		\end{cases}
\end{equation}
where $\Omega_{0,s}=\mathbf{u}_s^H\left(\sum_{d=1}^D\mathbf{W}_d+\mathbf{V}\right)\mathbf{u}_s+\sum_{u=1}^{U}p_u\left|G_{u,s}\right|^2+N_\mathrm{S}\sigma_{\mathrm{S},s}^{2}$ and $\Omega_{1,s}=(\mathbf{g}_s+\mathbf{u}_s)^H\left(\sum_{d=1}^D\mathbf{W}_d+\mathbf{V}\right)(\mathbf{g}_s+\mathbf{u}_s)+\sum_{u=1}^{U}p_u\left|G_{u,s}\right|^2+N_\mathrm{S}\sigma_{\mathrm{S},s}^{2}$. According to the Neyman-Pearson criterion, the maximum target detection probability for SE $s$ under an optimal threshold is derived as
\begin{equation}{P}^{\mathrm{SE}}_{\mathrm{D},s}=({\Omega_{1,s}}/{\Omega_{0,s}})^{-{\Omega_{0,s}}/(\Omega_{1,s}-\Omega_{0,s})}.\end{equation}

\vspace{-1em}
\subsection{Problem Formulation}
Our objective is to maximize the SSR, defined as the aggregate of non-negative secrecy rates achievable for all UL and DL LUs, which can be expressed as
$R_{\mathrm{SSR}}=\sum_{u=1}^{U}\left[R_{u}^{\mathrm{U}}-R_{u}^{\mathrm{EU}}\right]^{+}+\sum_{d=1}^{D}\left[R_{d}^{\mathrm{D}}-R_{d}^{\mathrm{ED}}\right]^{+}$,
which is achieved by jointly optimizing the receive combiners $\{\mathbf{u}_{i}\}$, transmit beamformers $\{\mathbf{W}_{d}\}$, AN covariance matrix $\mathbf{V}$, UL power allocation $\{p_{u}\}$, transmit and receive MA positions $\mathbf{t}^\mathrm{t}=\{\mathbf{t}_m^\mathrm{t}\}^{M_\mathrm{t}}_{m=1}$ and $\mathbf{r}^\mathrm{r}=\{\mathbf{r}_p^\mathrm{r}\}^{M_\mathrm{r}}_{p=1}$. 

Furthermore, each TBS $m$ operates under a maximum transmit power budget $P^{\mathrm{max}}_{\mathrm{D}}$. To facilitate the representation of per-TBS power consumption within the global beamforming formulation, we introduce an auxiliary selection matrix $\mathbf{A}_m=[\mathbf{0},\ldots,\mathbf{I}_{N_\mathrm{t}},\ldots,\mathbf{0}]\in\mathbb{C}^{N_\mathrm{t}\times M_\mathrm{t}N_\mathrm{t}}$, where $\mathbf{I}_{N_\mathrm{t}}$ occupies the $m$-th block while all other blocks are zero. Consequently, the transmit power constraint for TBS $m$ is expressed as
\begin{equation}
	\mathbb{E}\left(\mathbf{x}_m^H\mathbf{x}_m\right)\!=\!\operatorname{Tr}\left(\mathbf{A}_m\left({\sum}_{d=1}^D\mathbf{W}_d\!+\!\mathbf{V}\right)\mathbf{A}_m^H\right)\!\leq \! P^{\mathrm{max}}_{\mathrm{D}},\forall m.\label{TBS_power_constraint}
	\end{equation}
Based on the aforementioned system model, the joint optimization problem is formulated as follows 
\begin{subequations}\label{optimization problem P0}
	\begin{align}
		&\underset{\mathbf{u}_{i},\mathbf{W}_{d},\mathbf{V},p_{u},\mathbf{t}^\mathrm{t},\mathbf{r}^\mathrm{r}}{\max}~~R_{\mathrm{SSR}} \label{objective function P0} \\
		\text{s.t.}\quad
		&\eqref{TBS_power_constraint},\\
		&{R}^\mathrm{S}\geq\beta^\mathrm{S},\label{sensing MI}\\
		&{P}^{\mathrm{SE}}_{\mathrm{D},s}\leq\Upsilon^{\mathrm{SE}},\forall s,\label{SE probability}\\
		&0\leq p_{u}\leq P^{\mathrm{max}}_{\mathrm{U}},\forall u,\label{UL_power_constraint}\\
		&\mathbf{W}_d\succeq\mathbf{0},\operatorname{rank}(\mathbf{W}_d)\leq1,\forall d, \label{Wd_rank 1 P0}\\
		&\mathbf{V}\succeq\mathbf{0},\label{AN_constraint}\\
		&\mathbf{t}_m^\mathrm{t}\in\mathcal{C}^\mathrm{t}_{m},\mathbf{r}_p^\mathrm{r}\in\mathcal{C}^\mathrm{r}_{p},\forall m, p,\label{ranges of MA movements}\\
		&\left\|\mathbf{t}^\mathrm{t}_{m,a}\!-\!\mathbf{t}^\mathrm{t}_{m,\tilde{a}}\right\|_{2}\!\geq \! \breve{D},1\!\leq \!a \!\neq \!\tilde{a}\!\leq \!N_{\mathrm{t}},\forall m,\label{minimum TBS MA distance}\\
		&\left\|\mathbf{r}^\mathrm{r}_{p,b}\!-\!\mathbf{r}^\mathrm{r}_{p,\tilde{b}}\right\|_{2}\!\geq \! \breve{D},1\!\leq \! b \!\neq \!\tilde{b}\!\leq \!N_{\mathrm{r}},\forall p,\label{minimum RBS MA distance}
	\end{align}
\end{subequations}
where constraint \eqref{sensing MI} guarantees sensing performance by enforcing a minimum MI threshold $\beta^\mathrm{S}$, while \eqref{SE probability} ensures sensing security by limiting the maximum detection probability for each SE $s$ to a threshold $\Upsilon^{\mathrm{SE}}$. The UL transmission power for each LU $u$ is bounded by $P^{\mathrm{max}}_{\mathrm{U}}$ in \eqref{UL_power_constraint}. The constraints in \eqref{Wd_rank 1 P0} and \eqref{AN_constraint} impose the necessary semidefinite and rank-one properties on the transmit covariance matrices. Furthermore, constraint \eqref{ranges of MA movements} restricts the MAs to their designated regions $\mathcal{C}^\mathrm{t}_{m}$ and $\mathcal{C}^\mathrm{r}_{p}$. Finally, to satisfy practical hardware implementation requirements and prevent antenna coupling, constraints \eqref{minimum TBS MA distance} and \eqref{minimum RBS MA distance} enforce a minimum inter-MA spacing $\breve{D}$ at the TBSs and RBSs, respectively. 
Note that problem \eqref{optimization problem P0} is highly non-convex and difficult to solve. The complexity arises from the non-convex objective function \eqref{objective function P0}, the rank-one constraints, and the stringent minimum inter-MA distance constraints, compounded by the strong coupling between beamforming design, power allocation, and MA positions. As a result, directly obtaining a globally optimal solution for this problem is an extremely challenging task\footnote{In this work, perfect CSI for all links is assumed to be available to establish an ideal performance benchmark for the considered system. In practice, the CSI of LUs can be acquired through pilot-assisted channel estimation and feedback, while the channel parameters of the sensing target can be estimated through cooperative sensing and tracking \cite{10616023}. Effective channel estimation methods for MA systems have also been developed \cite{10318061,10659325}. However, accurate CSI of the eavesdropping links is generally difficult to obtain because CEs and SEs are passive and uncooperative. Perfect CSI of the eavesdropping links is adopted for idealized performance evaluation, while the eavesdroppers are modeled under a worst-case assumption with full channel knowledge and strong interception capabilities \cite{10818453}.
Section~\ref{4} further evaluates the proposed scheme under imperfect CSI of both CEs and SEs while retaining perfect CSI for the LUs and sensing target. A comprehensive robust design accounting for CSI uncertainty in the legitimate communication, sensing, and eavesdropping links is left for future work.}.

\vspace{-0.5em}
\section{Joint Beamforming Design and Antenna Position Optimization Framework}\label{3}

In this section, we propose an efficient AO algorithm to tackle the intractability of problem \eqref{optimization problem P0} and iteratively obtain a high-quality suboptimal solution. Exploiting the block-structure of the variables, the original highly coupled problem is decomposed into three tractable subproblems.
The proposed framework proceeds by iteratively updating the receive combiners $\{\mathbf{u}_{i}\}$, the transmit beamforming and power allocation variables $\{\{\mathbf{W}_{d}\}, \mathbf{V},\{p_{u}\}\}$,  and the antenna positions $\{\mathbf{t}^\mathrm{t},\mathbf{r}^\mathrm{r}\}$, in an alternating manner until convergence is achieved.

\vspace{-1.0em}
\subsection{Problem Reformulation}
Before proceeding with the iterative optimization, we first reformulate the constraints into more tractable forms. By defining $\gamma^\mathrm{S}=2^{\beta^\mathrm{S}}-1$, the non-linear sensing MI constraint in \eqref{sensing MI} is equivalently transformed into a linear sensing SINR constraint given by
\begin{equation}\text{SINR}_\mathrm{t}\geq\gamma^\mathrm{S}.\label{sensing_SINR}\end{equation} 

Next, we address the non-convex sensing eavesdropping probability constraint in \eqref{SE probability}. The detection probability is governed by the function $\mathrm{P}^{\mathrm{SE}}_{\mathrm{D},s}=({\Omega_{1,s}}/{\Omega_{0,s}})^{-{1}/({{\Omega_{1,s}}/\Omega_{0,s}-1})}$. 
We observe that the function $f(x)=x^{-{1}/{(x-1)}}$ is monotonically increasing for $x \in (0, 1) \cup (1, \infty)$. Let $\Gamma^{\mathrm{SE}}$ denote the solution to the equation $f(\Gamma^{\mathrm{SE}})=\Upsilon^{\mathrm{SE}}$. The constraint in \eqref{SE probability} is then equivalently rewritten as
\begin{equation}{\Omega_{1,s}}/{\Omega_{0,s}}\leq\Gamma^{\mathrm{SE}},\forall s.\label{SE_SINR}\end{equation}

\vspace{-1.0em}
\subsection{Receive Beamforming Optimization}
In this subproblem, we optimize the receive combiners $\{\mathbf{u}_{i}\}$ while keeping the antenna positions $\{\mathbf{t}^\mathrm{t},\mathbf{r}^\mathrm{r}\}$, transmit beamformers $\{\{\mathbf{W}_{d}\}, \mathbf{V}\}$, and UL power $\{p_{u}\}$ fixed. Note that the receive combiners employed at the RBSs do not influence the signal reception at the eavesdroppers. Therefore, optimizing the receive combiners can be decoupled into maximizing the corresponding sensing SINR and UL communication SINRs.
The corresponding optimization problems are formulated as
\begin{equation}
	\begin{aligned}\label{sensing_SINR_max}
		\underset{\mathbf{u}_{0}}{\operatorname*{\mathrm{max}}}\quad\text{SINR}_\mathrm{t},
	\end{aligned}
	\vspace{-0.5em}
\end{equation}
\begin{equation}
	\begin{aligned}\label{UL_SINR_max}
		\underset{\mathbf{u}_{u}}{\operatorname*{\mathrm{max}}}\quad\text{SINR}^\mathrm{U}_{u},\forall u.
	\end{aligned}
	\vspace{-0.5em}
\end{equation}
Both subproblems are generalized Rayleigh-quotient maximization problems and admit closed-form solutions that correspond to the minimum variance distortionless response (MVDR) beamformer.
Specifically, the optimal sensing combiner $\mathbf{u}_{0}^{\star}$ and UL receive combiners $\mathbf{u}_{u}^{\star}$ are given by
\begin{equation}\label{sensing_filter}
	\mathbf{u}_{0}^{\star}=\mathbf{Q}_{0}^{-1}\mathbf{g}^\mathrm{r}_{0},
\end{equation}
\begin{equation}\label{UL_filter}
	\mathbf{u}_{u}^{\star}=(\mathbf{Q}_{u}^\mathrm{U})^{-1}\mathbf{h}_{u},\forall u,
	\vspace{-0.5em}
\end{equation}
respectively, where we define $\mathbf{Q}_{0}=\sum_{c=1}^{K_\mathrm{C}}\mathbf{G}_c\mathbf{R}\mathbf{G}_c^{H}+\sum_{u=1}^{U}p_{u}\mathbf{h}_{u}\mathbf{h}_{u}^H+\sigma_\mathrm{r}^{2}\mathbf{I}\in\mathbb{C}^{M_\mathrm{r}N_\mathrm{r}\times M_\mathrm{r}N_\mathrm{r}}$ and $\mathbf{Q}_{u}^\mathrm{U}=\sum_{u'\neq u}^{U}p_{u'}\mathbf{h}_{u'}\mathbf{h}_{u'}^H+\sum_{c=0}^{K_\mathrm{C}}\mathbf{G}_{c}\mathbf{R}\mathbf{G}^{H}_{c}+\sigma_\mathrm{r}^{2}\mathbf{I}\in\mathbb{C}^{M_\mathrm{r}N_\mathrm{r}\times M_\mathrm{r}N_\mathrm{r}}$. $\mathbf{R}\triangleq\sum_{d=1}^D\mathbf{W}_{d}+\mathbf{V}$ represents the total transmit covariance matrix. It is worth noting that the SINR metric is invariant to the scaling of the receive combiners. Thus, the solutions in \eqref{sensing_filter} and \eqref{UL_filter} are optimal regardless of any positive scalar multiplication.

\begin{figure*}[hb] 
	\centering
	\vspace{-1.0em}
	\hrulefill 
	\setcounter{equation}{39}
	\begin{align}\label{function f}
		\begin{aligned}
			f_1&={\sum}_{u=1}^{U}\log_{2}\left({\sum}_{u'=1}^{U}p_{u'}|\tilde{h}^{\mathrm{U}}_{u,u'}|^{2}+{\sum}_{c=0}^{K_\mathrm{C}}{\sum}_{d=1}^{D}\mathrm{Tr}\left\{\tilde{\mathbf{G}}_{u,c}\mathbf{W}_{d}\right\}+{\sum}_{c=0}^{K_\mathrm{C}}\mathrm{Tr}\left\{\tilde{\mathbf{G}}_{u,c}\mathbf{V}\right\}+\sigma_\mathrm{r}^{2}\|\mathbf{u}_{u}\|_{2}^{2}\right)\\
			&+{\sum}_{d=1}^{D}\log_{2}\left({\sum}_{d'=1}^{D}\mathrm{Tr}\left\{\mathbf{H}_{d}\mathbf{W}_{d'}\right\}+\mathrm{Tr}\left\{\mathbf{H}_{d}\mathbf{V}\right\}+{\sum}_{u=1}^{U}p_{u}|h^{\mathrm{UD}}_{u,d}|^{2}+\sigma_{\mathrm{D},d}^{2}\right)+(U+D)\log_{2}\left({\prod}_{e=1}^{E}P_e\right),
		\end{aligned}
	\end{align} 
\end{figure*}

\vspace{-1.0em}
\subsection{Transmit Beamforming and Power Optimization}
Having determined the optimal receive combiners $\mathbf{u}_i^\star$, we proceed to optimize the transmit beamformers $\{\mathbf{W}_d\}$, AN covariance $\mathbf{V}$, and UL power $\{p_u\}$. Substituting $\mathbf{u}_{0}^{\star}$ from \eqref{sensing_filter} into the sensing SINR expression yields
\setcounter{equation}{34}
\begin{equation}\label{equivalent SINR_t}
	\widetilde{\text{SINR}}_\mathrm{t}=\sigma_{0}^{2}(\mathbf{g}^\mathrm{t}_{0})^{H}\mathbf{R}\mathbf{g}^\mathrm{t}_{0}(\mathbf{g}^\mathrm{r}_{0})^H\mathbf{Q}_{0}^{-1}\mathbf{g}^\mathrm{r}_{0}.
\end{equation}
Since the covariance matrices $\{\mathbf{W}_d \succeq \mathbf{0}, \forall d\}$ and $\mathbf{V} \succeq \mathbf{0}$ are positive semi-definite, it follows that $\mathbf{R} \succeq \mathbf{0}$. Furthermore, to satisfy the strictly positive sensing SINR threshold $\gamma^\mathrm{S} > 0$, the condition $(\mathbf{g}^\mathrm{t}_{0})^{H}\mathbf{R}\mathbf{g}^\mathrm{t}_{0} > 0$ must strictly hold. Therefore, the constraint \eqref{sensing_SINR} can be equivalently reformulated as
\begin{equation}\begin{aligned}\label{equivalent sensing constraint}
		(\mathbf{g}^\mathrm{r}_{0})^H\mathbf{Q}_\mathrm{0}^{-1}\mathbf{g}^\mathrm{r}_{0}
		&\geq\frac{\gamma^\mathrm{S}}{\sigma_{0}^{2}}((\mathbf{g}^\mathrm{t}_{0})^{H}\mathbf{R}\mathbf{g}^\mathrm{t}_{0})^{-1}.
\end{aligned}\end{equation}

The function $g(\mathbf{X})=\mathbf{g}^H\mathbf{X}^{-1}\mathbf{g}$ is known to be convex with respect to $\mathbf{X}$ for $\mathbf{X} \succ \mathbf{0}$. Since $\mathbf{Q}_{0}(\mathbf{W}_d,\mathbf{V})$ is affine in the optimization variables, the left-hand side (LHS) of \eqref{equivalent sensing constraint} is a convex function of $\{\{\mathbf{W}_{d}\},\mathbf{V},\{p_{u}\}\}$. Similarly, $((\mathbf{g}_{0}^\mathrm{t})^{H}\mathbf{R}\mathbf{g}^\mathrm{t}_{0})^{-1}$ on the right-hand side is convex with respect to $\{\{\mathbf{W}_{d}\},\mathbf{V}\}$. This structure identifies \eqref{equivalent sensing constraint} as a difference-of-convex (DC) constraint. To address this, we employ the SCA method \cite{11249714}. 
Specifically, at the $k_\mathrm{a}$-th iteration, the convex LHS is approximated by its first-order Taylor lower bound around $\mathbf{Q}_{0}^{(k_\mathrm{a}-1)}$. The resulting lower bound is given by
\begin{align}\label{sensing constraint upper bound}
	&(\mathbf{g}^\mathrm{r}_{0})^H\mathbf{Q}_\mathrm{0}^{-1}\mathbf{g}^\mathrm{r}_{0}
	\!\geq\!(\mathbf{g}^\mathrm{r}_{0})^H\left(\mathbf{Q}_\mathrm{0}^{(k_\mathrm{a}-1)}\right)^{-1}\mathbf{g}^\mathrm{r}_{0}\!-\!(\mathbf{g}^\mathrm{r}_{0})^H\left(\mathbf{Q}_\mathrm{0}^{(k_\mathrm{a}-1)}\right)^{-1}\nonumber\\
	&\times\!\!\left(\mathbf{Q}_\mathrm{0}\!-\!\mathbf{Q}_\mathrm{0}^{(k_\mathrm{a}-1)}\right)\!\!\left(\mathbf{Q}_\mathrm{0}^{(k_\mathrm{a}-1)}\right)^{-1}\mathbf{g}^\mathrm{r}_{0}
	\!\triangleq\! f^\mathrm{r}_{0}\left(\mathbf{Q}_\mathrm{0},\mathbf{Q}_\mathrm{0}^{(k_\mathrm{a}-1)}\right),
\end{align}
where $\mathbf{Q}_{0}^{(k_\mathrm{a}-1)}$ and $\mathbf{R}^{(k_\mathrm{a}-1)}$ are computed using the solution from the $(k_\mathrm{a}-1)$-th iteration. Replacing the LHS of \eqref{equivalent sensing constraint} with this approximation yields the convex constraint
\begin{equation}\label{sensing constraint transform}
	f^\mathrm{r}_{0}\left(\mathbf{Q}_\mathrm{0},\mathbf{Q}_\mathrm{0}^{(k_\mathrm{a}-1)}\right)\geq\frac{\gamma^\mathrm{S}}{\sigma_{0}^{2}}((\mathbf{g}^\mathrm{t}_{0})^{H}\mathbf{R}\mathbf{g}^\mathrm{t}_{0})^{-1}.
\end{equation}

Next, the sensing eavesdropping constraint in \eqref{SE_SINR} is reformulated as
\begin{align}\label{SE eavesdropping constraint transform}
	&(\mathbf{g}_s\!+\!\mathbf{u}_s)^H\mathbf{R}(\mathbf{g}_s+\mathbf{u}_s)+{\sum}_{u=1}^{U}p_u\left|G_{u,s}\right|^2+N_\mathrm{S}\sigma_{\mathrm{S},s}^{2} \nonumber\\
&\leq\Gamma^{\mathrm{SE}}\bigg(\mathbf{u}_s^H\mathbf{R}\mathbf{u}_s\!+\!{\sum}_{u=1}^{U}p_u\left|G_{u,s}\right|^2+N_\mathrm{S}\sigma_{\mathrm{S},s}^{2}\bigg),\forall s,
\end{align}
which is an affine constraint with respect to the optimization variables. Subsequently, we address the non-convex objective function. Using the properties of logarithmic functions, the SSR can be expressed as a difference of two concave functions \cite{10818453}, i.e., $R_\mathrm{SSR} = f_1 - f_2$, where $f_1$ is defined in \eqref{function f}, shown at the bottom of the page, and  $f_2$ is given by
\setcounter{equation}{40}  
\begin{align}\label{function g}
	\begin{aligned}
		f_2\!=\!\sum_{u=1}^{U}\left(\log_{2}A_u\!+\!\log_{2}B_u\right)\!+\!\sum_{d=1}^{D}\left(\log_{2}C_d\!+\!\log_{2}D_d\right),
	\end{aligned}
\end{align}
where we introduce $\tilde{\mathbf{G}}_{u,c}=\mathbf{G}_{c}^H\mathbf{u}_{u}\mathbf{u}_{u}^{H}\mathbf{G}_{c}$ and $\tilde{h}^{\mathrm{U}}_{u,u'}=\mathbf{u}_{u}^{H}\mathbf{h}_{u'}$ to simplify notation. Accordingly, the auxiliary terms are defined as $A_{u}=\sum_{u'\neq u}^{U}p_{u'}|\tilde{h}^{\mathrm{U}}_{u,u'}|^{2}+\sum_{c=0}^{K_\mathrm{C}}\sum_{d=1}^{D}\mathrm{Tr}\left\{\tilde{\mathbf{G}}_{u,c}\mathbf{W}_{d}\right\}+\sum_{c=0}^{K_\mathrm{C}}\mathrm{Tr}\left\{\tilde{\mathbf{G}}_{u,c}\mathbf{V}\right\}+\sigma_\mathrm{r}^{2}\|\mathbf{u}_{u}\|_{2}^{2}$, $C_{d}=\sum_{d'\neq d}^{D}\mathrm{Tr}\left\{\mathbf{H}_{d}\mathbf{W}_{d'}\right\}+\mathrm{Tr}\left\{\mathbf{H}_{d}\mathbf{V}\right\}+\sum_{u=1}^{U}p_{u}|h^{\mathrm{UD}}_{u,d}|^{2}+\sigma_{\mathrm{D},d}^{2}$, and  $P_e=\mathrm{Tr}\{\mathbf{H}_e\mathbf{V}\}+\sigma_{\mathrm{E},e}^2$.
Based on these, we further define $B_{u}=\sum_{e=1}^{E}\left(\left|h^{\mathrm{UE}}_{u,e}\right|^{2}p_{u}\prod_{e'\neq e}^{E}P_{e'}\right)
+\prod_{e=1}^{E}P_e$ and $D_{d}=\sum_{e=1}^{E}\left(\mathrm{Tr}\left\{\mathbf{H}_{e}\mathbf{W}_{d}\right\}\prod_{e'\neq e}^{E}P_{e'}\right)+
\prod_{e=1}^{E}P_e$.

Thus, the optimization subproblem is formulated as
\vspace{-0.5em}
\begin{subequations}\label{optimization problem P1}
	\begin{align}
		&\underset{\mathbf{W}_{d},\mathbf{V},p_{u}}{\max}~~f_1-f_2 \label{objective function P1} \\
		\text{s.t.}\quad
		&\eqref{TBS_power_constraint}, \eqref{UL_power_constraint},\eqref{Wd_rank 1 P0}, \eqref{AN_constraint},\eqref{sensing constraint transform},\eqref{SE eavesdropping constraint transform}.
	\end{align}
\end{subequations}
Problem \eqref{optimization problem P1} remains non-convex due to the objective and the rank-one constraints. To apply SCA, we construct a global affine upper bound for the concave function $f_2$ at the $k_\mathrm{a}$-th iteration point $(\mathbf{W}_{d}^{(k_\mathrm{a})},\mathbf{V}^{(k_\mathrm{a})},p_{u}^{(k_\mathrm{a})})$ using the first-order Taylor expansion, i.e., $f_2^{\mathrm{ub}}\left(\mathbf{W}_{d},\mathbf{V},p_{u}\left|\mathbf{W}_{d}^{(k_\mathrm{a})},\mathbf{V}^{(k_\mathrm{a})},p_{u}^{(k_\mathrm{a})}\right.\right)$, which is given by
\eqref{function g_ub}, shown at the bottom of the next page.  
\begin{figure*}[hb] 
	\centering
	\vspace{-1.2em}
	\hrulefill 
	\begin{align}\label{function g_ub}
			f_2\left(\mathbf{W}_{d},\mathbf{V},p_{u}\right)&\leq f_2\left(\mathbf{W}_{d}^{(k_\mathrm{a})},\mathbf{V}^{(k_\mathrm{a})},p_{u}^{(k_\mathrm{a})}\right)+{\sum}_{d=1}^D\mathrm{Tr}\left\{\left(\nabla_{\mathbf{W}_{d}}f_2\left(\mathbf{W}_{d}^{(k_\mathrm{a})},\mathbf{V}^{(k_\mathrm{a})},p_{u}^{(k_\mathrm{a})}\right)\right)^{H}\left(\mathbf{W}_{d}-\mathbf{W}_{d}^{(k_\mathrm{a})}\right)\right\}\nonumber\\
			&+\mathrm{Tr}\left\{\left(\nabla_{\mathbf{V}}f_2\left(\mathbf{W}_{d}^{(k_\mathrm{a})},\mathbf{V}^{(k_\mathrm{a})},p_{u}^{(k_\mathrm{a})}\right)\right)^{H}\left(\mathbf{V}-\mathbf{V}^{(k_\mathrm{a})}\right)\right\}+{\sum}_{u=1}^U\nabla_{p_{u}}f_2\left(\mathbf{W}_{d}^{(k_\mathrm{a})},\mathbf{V}^{(k_\mathrm{a})},p_{u}^{(k_\mathrm{a})}\right)\left(p_{u}-p_{u}^{(k_\mathrm{a})}\right)\nonumber\\
			&\triangleq f_2^{\mathrm{ub}}\left(\mathbf{W}_{d},\mathbf{V},p_{u}\left|\mathbf{W}_{d}^{(k_\mathrm{a})},\mathbf{V}^{(k_\mathrm{a})},p_{u}^{(k_\mathrm{a})}\right.\right),
	\end{align}
\end{figure*}
The gradients $\nabla_{\mathbf{W}_d}f_2$, $\nabla_{\mathbf{V}}f_2$, and $\nabla_{p_{u}}f_2$ are explicitly derived in \eqref{delta_W}, \eqref{delta_V}, and \eqref{delta_qu} on the next page, 
\begin{figure*}[hb] 
	\centering
	\vspace{-1.2em}
	\hrulefill 
	\begin{subequations} 
		\begin{align}
			&\begin{aligned}\label{delta_W}
				\nabla_{\mathbf{W}_{d}}f_2=\frac{1}{\ln2}\left({\sum}_{u=1}^U\frac{1}{A_u}{\sum}_{c=0}^{K_\mathrm{C}}\tilde{\mathbf{G}}_{u,c}^H+{\sum}_{d^{\prime}\neq d}^D\frac{1}{C_{d^{\prime}}}\mathbf{H}^H_{d^{\prime}}+\frac{1}{D_d}{\sum}_{e=1}^{E}\left(\mathbf{H}^H_e{\prod}_{e^{\prime}\neq e}^{E}P_{e'}\right)\right),
			\end{aligned}\\
			&\begin{aligned}\label{delta_V}
				\nabla_{\mathbf{V}}f_2&=\sum_{u=1}^U\frac{1}{A_u\ln2}\sum_{c=0}^{K_\mathrm{C}}\tilde{\mathbf{G}}^H_{u,c}+\sum_{u=1}^U\frac{1}{B_u\ln2}\left[\sum_{e=1}^{E}|h_{u,e}^{\mathrm{UE}}|^2p_u\sum_{e^{\prime}\neq e}\left(\mathbf{H}^H_{e^{\prime}}\prod^{E}_{e^{\prime\prime}\neq e,e^{\prime\prime}\neq e^{\prime}}P_{e^{\prime\prime}}\right)+\sum_{e=1}^{E}\left(\mathbf{H}^H_e\prod^{E}_{e^{\prime}\neq e}P_{e^{\prime}}\right)\right]\\&+\sum_{d=1}^D\frac{1}{C_d\ln2}\mathbf{H}^H_d+\sum_{d=1}^D\frac{1}{D_d\ln2}\left[\sum_{e=1}^{E}\mathrm{Tr}\{\mathbf{H}_e\mathbf{W}_d\}\sum^{E}_{e^{\prime}\neq e}\left(\mathbf{H}^H_{e^{\prime}}\prod^{E}_{e^{\prime\prime}\neq e,e^{\prime\prime}\neq e^{\prime}}P_{e^{\prime\prime}}\right)+\sum_{e=1}^{E}\left(\mathbf{H}^H_e\prod^{E}_{e^{\prime}\neq e}P_{e^{\prime}}\right)\right],
			\end{aligned}\\
			&\begin{aligned}\label{delta_qu}
				\nabla_{p_{u}}f_2=\frac{1}{\ln2}\left({\sum}_{u^{\prime}\neq u}^{U}\frac{1}{A_{u^{\prime}}}|\tilde{h}_{u,u'}|^{2}+\frac{1}{B_{u}}{\sum}_{e=1}^{{E}}\left(|h_{u,e}^{\mathrm{UE}}|^{2}{\prod}^{E}_{e^{\prime}\neq e}P_{e^{\prime}}\right)+{\sum}_{d=1}^{D}\frac{1}{C_{d}}|h_{u,d}^{\mathrm{UD}}|^{2}\right),
			\end{aligned}
		\end{align}
	\end{subequations} 
\end{figure*}

By dropping the rank-one constraints \eqref{Wd_rank 1 P0}, a convex relaxation of the original problem at the $k_\mathrm{a}$-th iteration can be given as 
\begin{subequations}\label{optimization problem P2}
\begin{align}
		&\underset{\mathbf{W}_{d},\mathbf{V},p_{u}}{\max}~~f_1-f_2^{\mathrm{ub}} \label{objective function P2} \\
		\text{s.t.}\quad
		&\eqref{TBS_power_constraint}, \eqref{UL_power_constraint}, \eqref{AN_constraint},\eqref{sensing constraint transform},\eqref{SE eavesdropping constraint transform},\mathbf{W}_d\succeq\mathbf{0},\forall d.
\end{align}
\end{subequations}
This relaxed problem is convex and can be efficiently solved using standard solvers such as CVX. Although the relaxation yields a general rank solution, we can construct a rank-one optimal solution satisfying the original constraints without loss of optimality. 
Specifically, given the relaxed optimal solution $(\{\overline{\mathbf{W}}_d\},\overline{\mathbf{V}},\{\overline{p}_u\})$, we construct the rank-one beamformers as
\begin{equation}\mathbf{W}_d^{\star}=\mathbf{w}_{d}^{\star}(\mathbf{w}_{d}^{\star})^H,\forall d,\label{Wd_rank1}\end{equation}
where $\mathbf{w}_d^{\star}=\left(\mathbf{h}_{d}^{H}\overline{\mathbf{W}}_{d}\mathbf{h}_{d}\right)^{-1/2}\overline{\mathbf{W}}_{d}\mathbf{h}_{d},\forall d$.

Accordingly, the AN covariance matrix is updated as 
\begin{equation}\mathbf{V}^{\star}={\sum}_{d=1}^{D}\overline{\mathbf{W}}_{d}+\overline{\mathbf{V}}-{\sum}_{d=1}^{D}\mathbf{w}_{d}^{\star}(\mathbf{w}_{d}^{\star})^{H},\label{AN_rank1}\end{equation}
while maintaining the same power allocation 
\begin{equation}p_u^{\star}=\overline{p}_{u},\forall u.\label{UL_power_rank1}\end{equation}
This construction preserves the objective value of the relaxed solution and satisfies all constraints, including the power budget and rank-one requirements.

\vspace{-1em}
\subsection{Transmit and Receive Antenna Position Optimization}
Given the optimized transceiver beamforming $\{\{\mathbf{W}_{d}\},\mathbf{V}\}$, receive filters $\{\mathbf{u}_{i}\}$, and UL power allocation $\{p_{u}\}$, the SSR becomes solely a function of $\mathbf{t}^\mathrm{t}$ and $\mathbf{r}^\mathrm{r}$. Consequently, the antenna position optimization subproblem is formulated as
\begin{subequations}\label{optimization problem 10}
	\begin{align}
		&\underset{\mathbf{t}^\mathrm{t},\mathbf{r}^\mathrm{r}}{\max}~~R_{\mathrm{SSR}}(\mathbf{t}^\mathrm{t},\mathbf{r}^\mathrm{r}) \label{objective function 1} \\
		\text{s.t.}\quad
		&\eqref{ranges of MA movements},\eqref{minimum TBS MA distance},\eqref{minimum RBS MA distance},\eqref{sensing_SINR},\eqref{SE_SINR}.
	\end{align}
\end{subequations}

To address the inherent non-convexity and high dimensionality of problem \eqref{optimization problem 10}, we propose a SA-MCPSO algorithm. Distinct from standard PSO, which relies on a single velocity update, the SA-MCPSO generates multiple candidate velocities for each particle, to simultaneously optimize the positions of all MAs at the TBSs and RBSs.
This mechanism enables simultaneous exploration of the global search space and exploitation of local regions, thereby achieving a better balance between global exploration and local refinement.

The algorithm commences by initializing a swarm of $Q$ particles, with positions and velocities represented by the matrices $\mathbf{P}^{(0)}\!\!=\!\!\left\{\mathbf{p}_{1}^{(0)},\ldots,\mathbf{p}_{Q}^{(0)}\right\}$ and $\mathbf{Z}^{(0)}\!=\!\left\{\mathbf{z}_{1}^{(0)},\ldots,\mathbf{z}_{Q}^{(0)}\right\}$, where $\mathbf{P}^{(0)}, \mathbf{Z}^{(0)}\!\in\! \mathbb{R}^{2(M_{\mathrm{t}}N_{\mathrm{t}}+M_{\mathrm{r}}N_{\mathrm{r}})\times Q}$. Specifically, the position vector of the $q$-th particle, $\mathbf{p}_q^{(0)}$, encapsulates the complete configuration of all MAs, structured as $\mathbf{p}_q^{(0)}\!=\!\left[(\mathbf{t}_{1,q}^{\mathrm{t},(0)})^T,\dots,(\mathbf{t}_{M_{\mathrm{t}},q}^{\mathrm{t},(0)})^T,(\mathbf{r}_{1,q}^{\mathrm{r},(0)})^T,\dots,(\mathbf{r}_{M_{\mathrm{r}},q}^{\mathrm{r},(0)})^T\right]^T$, where $\mathbf{t}_{m,q}^{\mathrm{t},(0)}$ and $\mathbf{r}_{p,q}^{\mathrm{r},(0)}$ denote the initial coordinate vectors for the transmit and receive MAs at TBSs and RBSs, respectively. 

Assuming each MA operates within a square region of size $A \times A$, the elements of $\mathbf{P}^{(0)}$ are initialized following a uniform distribution over  $[-A/2, A/2]$ to satisfy the MA movement region constraints.
The personal best position for the $q$-th particle, $\mathbf{p}_{\mathrm{pbest},q}$, is initialized as its current position $\mathbf{p}_q^{(0)}$, while the global best position $\mathbf{p}_{\mathrm{gbest}}$ is determined by the particle yielding the maximum fitness value among the initial population. 

To evaluate the quality of each particle while discouraging constraint violations, we define a penalty-based fitness function $\mathcal{F}(\cdot)$, which can be given as
\begin{equation}\mathcal{F}\left(\mathbf{p}_q^{(c)}\right)=R_{\mathrm{SSR}}\left(\mathbf{p}_q^{(c)}\right)-\mathcal{P}\left(\mathbf{p}_{q}^{(c)}\right),1\leq c\leq K_\mathrm{PSO}^{\mathrm{max}},\end{equation}
where $\mathbf{p}_q^{(c)}$ denotes the position of the $q$-th particle in the $c$-th iteration. 
Note that calculating $R_{\mathrm{SSR}}$ entails reconstructing the channel response matrices based on the updated antenna coordinates $\mathbf{p}_q^{(c)}$. The penalty function $\mathcal{P}\left(\mathbf{p}_{q}^{(c)}\right)$ aggregates violations of the minimum inter-MA spacing, sensing SINR, and sensing eavesdropping probability constraints, defined as:
\begin{equation}\mathcal{P}\left(\mathbf{p}_{q}^{(c)}\right)={\sum}_{l=1}^4g_{l}\left(\mathbf{p}_q^{(c)}\right),\end{equation}
where the individual penalty terms are defined as
\begin{subequations}
	\begin{align}
		&\!\!g_1\!\left(\!\mathbf{p}_{q}^{(c)}\!\right)\!\!=\!\sum_{m=1}^{M_\mathrm{t}} \!\tau_{\mathrm{t},m} \!\!\sum_{a=1}^{N_\mathrm{t}\!-\!1} \!\sum_{\tilde{a}=a+1}^{N_\mathrm{t}} \!\!\delta\!\left( \left\| \mathbf{t}_{m,a}^{\mathrm{t},(c)} \!-\! \mathbf{t}_{m,\tilde{a}}^{\mathrm{t},(c)} \right\|_2 \!<\! \breve{D} \right)\!\!, \\
		&\!\!g_2\!\left(\mathbf{p}_{q}^{(c)}\right)\!\!=\!
		\sum_{p=1}^{M_\mathrm{r}}\!\tau_{\mathrm{r},p} \sum_{b=1}^{N_\mathrm{r}\!-\!1}\!\sum_{\tilde{b}=b+1}^{N_\mathrm{r}}\!\!\delta\!\left(\left\|\mathbf{r}_{p,b}^{\mathrm{r},(c)}\!-\!\mathbf{r}_{p,\tilde{b}}^{\mathrm{r},(c)}\right\|_2\!<\!\breve{D}\right)\!,\\
		&\!\!g_3\left(\mathbf{p}_{q}^{(c)}\right)= \tau^{\mathrm{S}}\delta\left(\text{SINR}_\mathrm{t}<\gamma^\mathrm{S}\right),\\
		&\!\!g_4\left(\mathbf{p}_{q}^{(c)}\right)={\sum}_{s=1}^{S}\tau_{s}^{\mathrm{SE}}\delta\left({\Omega_{1,s}}/{\Omega_{0,s}}>\Gamma^{\mathrm{SE}}\right),
	\end{align}
\end{subequations}
where $\delta(\cdot)$ is an indicator function returning 1 if the corresponding constraint is violated and 0 otherwise. 
The penalty factors $\tau_{\mathrm{t},m}$, $\tau_{\mathrm{r},p}$, $\tau^{\mathrm{S}}$, and $\tau_{s}^{\mathrm{SE}}$ are set sufficiently large to impose substantial penalties on infeasible solutions while maintaining a proper trade-off between constraint enforcement and search stability. Although the indicator-based penalty introduces a non-smooth fitness landscape, it is suitable for the heuristic PSO framework since no gradient information is required. The penalty terms guide the swarm toward feasible regions during the search. Since the penalty-based optimization does not guarantee strict feasibility throughout the iterations, all final candidates are explicitly verified against the original constraints after convergence, and the best feasible candidate is selected as the final output.

In the $c$-th iteration, the SA-MCPSO generates $\overline{M}$ candidate velocities for the $q$-th particle. These candidates are synthesized from weighted combinations of $\overline{N}$ fundamental velocity components \cite{10818453}.  Specifically, the $\overline{m}$-th candidate velocity is given by $\mathbf{z}_{q,\overline{m}}^{(c)} = \boldsymbol{\Lambda}_q^{(c)} \mathbf{c}_{q,\overline{m}}$, where $\mathbf{c}_{q,\overline{m}}$ is a combination weight vector and $\boldsymbol{\Lambda}_q^{(c)}\!\!=\!\!\left[\boldsymbol{\lambda}_{q,1}^{(c)},\ldots,\boldsymbol{\lambda}_{q,\overline{N}}^{(c)}\right]$ collects the velocity components. The $\overline{n}$-th velocity component is calculated as
\begin{equation}\begin{aligned}\boldsymbol{\lambda}_{q,\overline{n}}^{(c)}&=\nu_{\overline{n}}\mathbf{z}_{q}^{(c-1)}+c_{1}\mathbf{e}_{1}\odot\left(\mathbf{p}_{\mathrm{pbest},q}-\mathbf{p}_{q}^{(c-1)}\right)\\&+c_2\mathbf{e}_2\odot\left(\mathbf{p}_{\mathrm{gbest}}-\mathbf{p}_q^{(c-1)}\right),\end{aligned}\end{equation}
where $\mathbf{z}_{q}^{(c-1)}$ represents the actual velocity selected in the previous iteration, $\nu_{\overline{n}}$ represents the inertia weight, and $c_1$ and $c_2$ are the cognitive and social learning factors, respectively. The random vectors $\mathbf{e}_1, \mathbf{e}_2 \in [0, 1]^{2(M_{\mathrm{t}}N_{\mathrm{t}}+M_{\mathrm{r}}N_{\mathrm{r}})\times1}$ introduce stochasticity to prevent premature convergence. 

This multi-candidate mechanism allows dynamic balancing of exploration and exploitation by varying $\mathbf{c}_{q,\overline{m}}$. 
Subsequently, $\overline{M}$ candidate positions are generated via 
\begin{equation}
	\mathbf{p}_{q,\overline{m}}^{(c)} = \mathcal{B} \left\{ \mathbf{p}_q^{(c-1)} + \mathbf{z}_{q,\overline{m}}^{(c)} \right\}, \quad \overline{m} = 1, \ldots, \overline{M},
\end{equation}
where the operator $\mathcal{B}\{\cdot\}$ projects particles exceeding the boundary $[-A/2, A/2]$ back to the feasible region,  which can be given by
\begin{equation}\left[\mathcal{B}\left\{\mathbf{p}\right\}\right]_i=\left\{
	\begin{array}{rcl}
		-{A}/{2},&\mathrm{if}&\left[\mathbf{p}\right]_i<-{A}/{2},\\
		{A}/{2},&\mathrm{if}&\left[\mathbf{p}\right]_i>{A}/{2},\\
		\left[\mathbf{p}\right]_i, &\mathrm{otherwise}.
	\end{array}\right.
\end{equation}	

Then, a greedy selection strategy is employed to determine the optimal candidate position update that maximizes the fitness function:
\begin{equation}\mathbf{p}_{q}^{(c)}={\arg\max}_{\mathbf{p}_{q,\overline{m}}^{(c)}}\left\{\mathcal{F}\left(\mathbf{p}_{q,1}^{(c)}\right),\ldots,\mathcal{F}\left(\mathbf{p}_{q,\overline{M}}^{(c)}\right)\right\}.\end{equation}
The particle's velocity for the next iteration is effectively updated to the specific candidate velocity $\mathbf{z}_{q,\overline{m}}^{(c)}$ corresponding to the selected position.  Following the position update, the personal best $\mathbf{p}_{\mathrm{pbest},q}$ and global best $\mathbf{p}_{\mathrm{gbest}}$ are updated if superior fitness values are achieved.

Upon completion of $K_\mathrm{PSO}^{\mathrm{max}}$  iterations, a post-processing step is implemented to minimize the total MA movement distance during reconfiguration. Specifically, we employ an antenna position matching algorithm to establish an optimal one-to-one mapping between the initial and optimized antenna coordinates. This procedure extends the methodology of Algorithm 4 in \cite{11075964} to the considered multi-static scenario, thereby facilitating practical implementation.

\vspace{-1em}
\subsection{Convergence and Complexity Analysis}
In this section, our proposed algorithm for the networked secure ISAC system is summarized in \textbf{Algorithm \ref{overall algorithm}}.
\begin{algorithm}[t!]
    \caption{Joint beamforming design and antenna position optimization algorithm.}
	\renewcommand{\algorithmicrequire}{\textbf{Input:}}
	\renewcommand{\algorithmicensure}{\textbf{Output:}}
	\begin{algorithmic}[1]
		\REQUIRE   SCA convergence threshold $\xi_{\mathrm{SCA}}$, AO convergence threshold $\xi_{\mathrm{AO}}$, and maximum iterations $K_\mathrm{AO}^{\mathrm{max}}$.
		\STATE \textbf{Initialize} iteration number $k=1$, and feasible variables $\{\mathbf{W}_{d}^{(0)},\forall d\}$,  $\mathbf{V}^{(0)}$, $\{p_{u}^{(0)},\forall u\}$, ${\mathbf{t}}^{\mathrm{t},(0)}$, and ${\mathbf{r}}^{\mathrm{r},(0)}$;
		\REPEAT
		\STATE  Update the optimal receive filters $\{\mathbf{u}_{i}^{(k)},\forall i\}$ via \eqref{sensing_filter} and \eqref{UL_filter};
		\STATE Given $\{\mathbf{u}_{i}^{(k)},\forall i\}$, ${\mathbf{t}}^{\mathrm{t},(k-1)}$ and ${\mathbf{r}}^{\mathrm{r},(k-1)}$, update $\{\mathbf{W}_{d}^{(k)},\forall d\}$,  $\mathbf{V}^{(k)}$ and  $\{p_{u}^{(k)},\forall u\}$ by 
		solving problem \eqref{optimization problem P2}  via the SCA algorithm, and recovering the rank-one solutions via \eqref{Wd_rank1}, \eqref{AN_rank1}, and \eqref{UL_power_rank1};
		\STATE Given $\{\mathbf{u}_{i}^{(k)},\forall i\}$, $\{\mathbf{W}_{d}^{(k)},\forall d\}$,  $\mathbf{V}^{(k)}$ and $\{p_{u}^{(k)},\forall u\}$, update  ${\mathbf{t}}^{\mathrm{t},(k)}$ and ${\mathbf{r}}^{\mathrm{r},(k)}$ via the SA-MCPSO algorithm;
		\STATE Update iteration index $k = k + 1$;
		\UNTIL The fractional diminution of the objective function value falls below the threshold $\xi_{\mathrm{AO}}$  or $k > K_\mathrm{AO}^{\mathrm{max}}$;
		\STATE Perform antenna position matching to minimize the total MA movement distance;
		\ENSURE $\{\mathbf{W}_{d}^{\star},\forall d\}$,  $\mathbf{V}^{\star}$,  $\{p_{u}^{\star},\forall u\}$, $\{\mathbf{u}_{i}^{\star},\forall i\}$, ${\mathbf{t}}^{\mathrm{t},\star}$, ${\mathbf{r}}^{\mathrm{r},\star}$.
	\end{algorithmic}\label{overall algorithm}  
\end{algorithm}

\subsubsection{Convergence Analysis}\label{Convergence Analysis}
The convergence of the overall AO algorithm is analyzed as follows. Let $k$ ($1\leq k\leq K_\mathrm{AO}^{\mathrm{max}}$) denote the current iteration index. The monotonic ascent of the SSR across alternating updates is guaranteed by the inequality sequence
\begin{align}
		&R_{\mathrm{SSR}}\left(\mathbf{u}_{i}^{(k-1)}, \mathbf{W}_{d}^{(k-1)},\mathbf{V}^{(k-1)},p_{u}^{(k-1)}, {\mathbf{t}}^{\mathrm{t},(k-1)},{\mathbf{r}}^{\mathrm{r},(k-1)}\right)\nonumber\\
		&\overset{(a)}{\operatorname*{\leq}}R_{\mathrm{SSR}}\left(\mathbf{u}_{i}^{(k)},\mathbf{W}_{d}^{(k-1)},\mathbf{V}^{(k-1)},p_{u}^{(k-1)},{\mathbf{t}}^{\mathrm{t},(k-1)},{\mathbf{r}}^{\mathrm{r},(k-1)}\right)\nonumber\\
		&\overset{(b)}{\operatorname*{\leq}}R_{\mathrm{SSR}}\left(\mathbf{u}_{i}^{(k)},\mathbf{W}_{d}^{(k)},\mathbf{V}^{(k)},p_{u}^{(k)},{\mathbf{t}}^{\mathrm{t},(k-1)},{\mathbf{r}}^{\mathrm{r},(k-1)}\right)\nonumber\\
		&\overset{(c)}{\operatorname*{\leq}}R_{\mathrm{SSR}}\left(\mathbf{u}_{i}^{(k)},\mathbf{W}_{d}^{(k)},\mathbf{V}^{(k)},p_{u}^{(k)},{\mathbf{t}}^{\mathrm{t},(k)},{\mathbf{r}}^{\mathrm{r},(k)}\right),
\end{align}
where inequality $(a)$ holds due to the optimal receive beamformer update in Step 3. Similarly, $(b)$ follows from the monotonic improvement property of the SCA-based transmit beamforming and power allocation update in Step 4, where the constructed surrogate problem is solved to obtain a non-decreasing objective value. Inequality $(c)$ follows from the greedy update mechanism of the SA-MCPSO algorithm in Step 5, where a new MA configuration is accepted only when it provides an improved feasible SSR value.
The convergence of the SA-MCPSO algorithm is facilitated by the bounded MA search regions and the greedy update mechanism, where the global best position is updated only when an improved feasible candidate is obtained. Therefore, the SSR sequence generated by the proposed AO framework is monotonically non-decreasing. Moreover, due to the finite transmit power constraints, the achievable SSR is upper-bounded. According to the monotone convergence theorem, the objective sequence converges to a finite value.
The iterative process terminates when the increment in $R_{\mathrm{SSR}}$ falls below a predefined threshold $\xi_{\mathrm{AO}}$ or the maximum number of iterations is reached. The convergence behavior is further verified by simulation results in Section~\ref{4}.

\subsubsection{Complexity Analysis}
The computational complexity of the proposed algorithm arises from three subproblems and final antenna position matching. First, receive filter generation scales as $\mathcal{O}((U+1)(M_\mathrm{r}N_\mathrm{r})^{3})$ due to matrix inversions. Second, transmit beamforming via SCA incurs a complexity of ${o}_\mathrm{a}=\mathcal{O}(I_{{\mathrm{a}}} ((D+1)(M_\mathrm{t}N_\mathrm{t})^{2}+U)^{3.5})$, where $I_\mathrm{a}$ denotes the SCA iterations. Third, for SA-MCPSO-based position optimization, the cost per particle update comprises channel reconstruction $o_{\mathrm{b},1}$ and fitness evaluation $o_{\mathrm{b},2}$, given by ${o}_{\mathrm{b},1} = \mathcal{O}( M_\mathrm{t} N_\mathrm{t} ( \sum_{d=1}^{D} L^\mathrm{t}_{m,d} + \sum_{e=1}^{E} L^\mathrm{t}_{m,e} + K_\mathrm{C}+1) + M_\mathrm{r}N_\mathrm{r}( \sum_{u=1}^{U} L^\mathrm{r}_{p,u} + K_\mathrm{C}+1))$ and ${o}_{\mathrm{b},2} = \mathcal{O}( M_\mathrm{t} \binom{N_\mathrm{t}}{2} + M_\mathrm{r} \binom{N_\mathrm{r}}{2} + (D+U+E+S)(M_\mathrm{t}N_\mathrm{t})^2 )$, respectively. With $Q$ particles, $I_\mathrm{b}$ iterations, and $\overline M$ candidate velocities, the SA-MCPSO complexity is ${o}_{\mathrm{b}} = \mathcal{O}(  I_\mathrm{b} Q \overline M (o_{\mathrm{b},1} + o_{\mathrm{b},2}) )$. Consequently, with $I_{\mathrm{AO}}$ outer iterations, the total AO complexity is ${o}_\mathrm{AO} = \mathcal{O}( I_{\mathrm{AO}} ( (U+1)(M_\mathrm{r}N_\mathrm{r})^{3} + o_{\mathrm{a}} + o_{\mathrm{b}} ) )$. Incorporating the antenna matching step with cost $o_{\mathrm{c}}=\mathcal{O}(\frac{1}{2}(M_{\mathrm{t}}N_{\mathrm{t}}(N_{\mathrm{t}}+1)+M_{\mathrm{r}}N_{\mathrm{r}}(N_{\mathrm{r}}+1)))$, the overall complexity is $\mathcal{O}(o_\mathrm{AO}+o_{{\mathrm{c}}})$.

\vspace{-1em}
\section{Simulation Results}\label{4}

\begin{table}[!t]
	\caption{Simulation Parameters}
	\vspace{-0.5em}
	\label{tab1}
	\centering
	\begin{tabular}{|l|l|}
		\hline
		\multicolumn{1}{|c|}{\textbf{Parameters}}  & \multicolumn{1}{c|}{\textbf{Values}} \\ \hline
		 Numbers of nodes $M_\mathrm{t}$, $M_\mathrm{r}$, $D$, $U$, ${E}$, ${S}$, ${K_\mathrm{C}}$ & 2  \\ \hline
	     Number of  SE antennas $N_\mathrm{S}$ & 4 \\ \hline
		 Moving region size $A \times A$ & $4\lambda \times 4\lambda$ \\ \hline
		 Minimum inter-MA distance $\breve{D}$ & $\lambda/2$ \\ \hline
         Path loss at the reference distance (dB) $\beta_0$ & -30   \\ \hline
		 Path loss exponent $\alpha$     & 2.5   \\ \hline
		 Average noise powers  (dBm) $\sigma_{\mathrm{r}}^2$, $\sigma _{\mathrm{D},d}^2$, $\sigma _{\mathrm{E}, e}^2$, $\sigma _{\mathrm{S}, s}^2$   & -90   \\ \hline
		 Maximum transmit power (dBm) $P^{\mathrm{max}}_{\mathrm{D}}$, $P^{\mathrm{max}}_{\mathrm{U}}$  & 40,10    \\ \hline
		 Sensing MI threshold (bps/Hz) 	$\beta^\mathrm{S}$  & 3.5    \\ \hline
		 Number of particles $Q$  & 100   \\ \hline
		 Maximum numbers of SA-MCPSO iterations $K_\mathrm{PSO}^{\mathrm{max}}$  & 100   \\ \hline
		 Maximum numbers of SCA/AO iterations $K_\mathrm{SCA}^{\mathrm{max}}$, $K_\mathrm{AO}^{\mathrm{max}}$  & 50   \\ \hline
		 Penalty factors $\tau_{\mathrm{t},m}$, $\tau_{\mathrm{r},p}$, $\tau^{\mathrm{S}}$, $\tau_{s}^{\mathrm{SE}}$  & 10   \\ \hline
		 Personal and global learning factors $c_1$, $c_2$  & 1.4   \\ \hline
		 Convergence thresholds $\xi_\mathrm{SCA}$, $\xi_\mathrm{AO}$  & $10^{-4}$   \\ \hline
	\end{tabular}
	\vspace{-1.5em}
\end{table}

In this section, we provide numerical results to evaluate the performance of the proposed scheme in a global three-dimensional (3D) Cartesian coordinate system.
The TBSs are located at $(0, 50, 20)$ m and $(50, 100, 20)$ m, while the RBSs are at $(50, 0, 20)$ m and $(100, 50, 20)$ m. The sensing target is positioned at $(50, 50, 20)$ m.
The DL LUs, UL LUs, CEs, and SEs are randomly and uniformly distributed on the ground within the coverage area of the BSs. Meanwhile, clutter scatterers are randomly distributed within a cubic region of side length $20$ m centered on the target.
Each TBS and RBS is equipped with $N_\mathrm{t}=N_\mathrm{r}\triangleq N=6$ MAs. We adopt a geometric multipath channel model with $L_{m,d}^\mathrm{t} = L_{m,e}^\mathrm{t} = L_{p,u}^\mathrm{r}\triangleq L=6$ propagation paths. 
The channels for TBS-DL LU, TBS-CE, and UL LU-RBS links comprise one LoS path and $L-1$ NLoS paths. Accordingly, the elements of PRVs $\mathbf{\Psi}$ are independently distributed as $\Psi_1 \sim \mathcal{CN}\left(0, \beta_0 d^{-\alpha} \kappa / (\kappa + 1)\right)$ for the LoS path, and $\Psi_l \sim \mathcal{CN}\left(0, \beta_0 d^{-\alpha} / ((\kappa + 1)(L - 1))\right)$, $l \geq 2$ for NLoS paths \cite{10944486},  where $\kappa$ denotes the Rician factor. Furthermore, the propagation coefficients $\beta_{m,c}^\mathrm{t}$, $\beta_{p,c}^\mathrm{r}$, and $\beta_{s,c}^\mathrm{S}$ are uniformly modeled as $\sqrt{\beta_0(d/d_0)^{-\alpha}}$, where $d$ is the corresponding propagation distance, $\beta_0 = -30$ dB represents the reference path loss at $d_0=1$ m, and $\alpha$ denotes the path loss exponent \cite{10694747}. The scattering coefficients $\alpha_c\sim\mathcal{CN}(0,\sigma_c^2)$ and $\alpha_{s,m,c}\sim\mathcal{CN}(0,\sigma_{s,m,c}^2)$ follow the Swerling-I model \cite{11203895}, where $\sigma_c^2$ and $\sigma_{s,m,c}^2$ denote the corresponding average scattering powers. 
Regarding security constraints, the sensing eavesdropping detection probability threshold is $\Upsilon^{\mathrm{SE}} = 0.4$ \cite{10605793}. 
The proposed SA-MCPSO algorithm employs $\overline M=3$ candidate velocities synthesized from $\overline N=3$ distinct velocity components. To balance exploration and exploitation, the inertia weights are set to $\nu_1\! = \!0.9$, $\nu_2 \!=\! 0.75$, and $\nu_3 \!= \! 0.4$ \cite{10818453}. The particle combination weight vectors are initialized as $\mathbf{c}_{q,1}\!=\!\left[ 1,0,0\right]^T $, $\mathbf{c}_{q,2}\!=\!\left[ 0,1,0\right]^T $ and $\mathbf{c}_{q,3}\!=\!\left[ 0,0,1\right]^T$. Unless otherwise specified, the default simulation parameters are summarized in Table~\ref{tab1}.

\begin{figure}[t!]
	\centering
	\includegraphics[height=45.7mm]{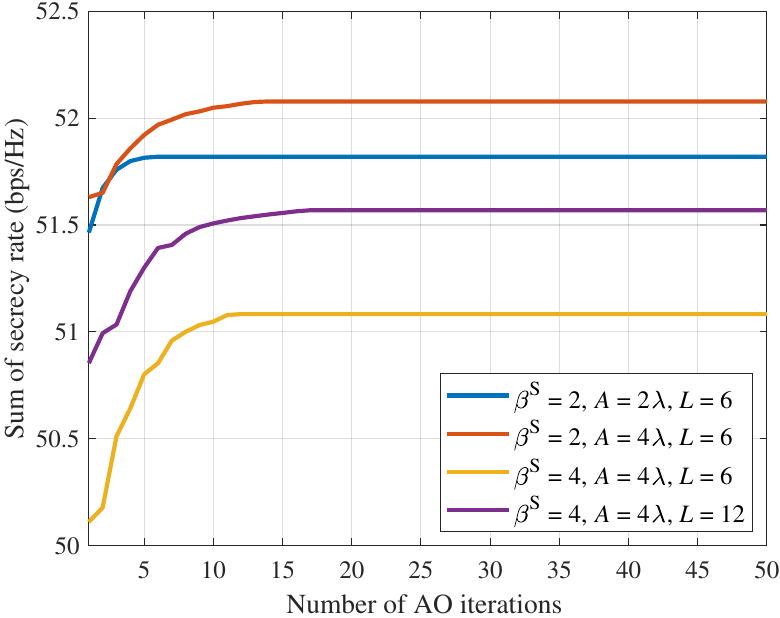}
	\vspace{-0.35em}
	\caption{Convergence evaluations of the proposed AO algorithm.}
	\vspace{-1em}
	\label{AO}
\end{figure}

The convergence behavior of the proposed AO algorithm is evaluated in Fig.~\ref{AO}. Consistent with the theoretical analysis in Section~\ref{Convergence Analysis}, the SSR increases monotonically and stabilizes within approximately 20 iterations under all considered configurations, verifying the efficiency of the proposed algorithm. Based on the converged results, we further analyze the impact of key parameters. First, increasing the number of propagation paths from $L=6$ to $L=12$ improves the SSR by approximately $0.5$ bps/Hz. This improvement can be attributed to the richer multipath structure, which enhances spatial channel diversity. Second, enlarging the normalized moving region $A/\lambda$ from 2 to 4 yields a higher secrecy rate. The enlarged region provides greater flexibility for MA positioning within the feasible spatial domain. Conversely, increasing the sensing MI threshold $\beta^\mathrm{S}$ from 2 to 4 bps/Hz reduces the achievable SSR. This reflects the fundamental communication-sensing trade-off, as more stringent sensing requirements restrict the feasible resource allocation strategies and limit the achievable communication secrecy performance.
\begin{figure*}[t]
  \centering
  \subfloat[Communication-only signals]{
   \includegraphics[width=0.255\textwidth]{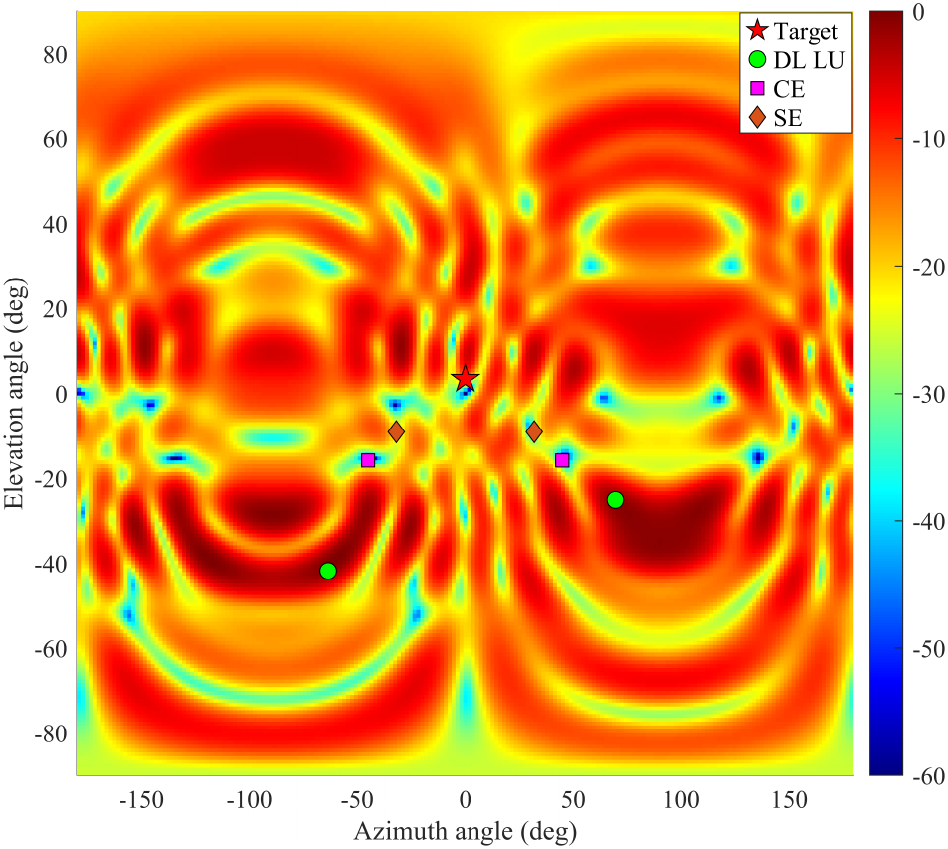}
   \label{com_only}
  }
  \hfill  
  \subfloat[AN-only signals]{
    \includegraphics[width=0.255\textwidth]{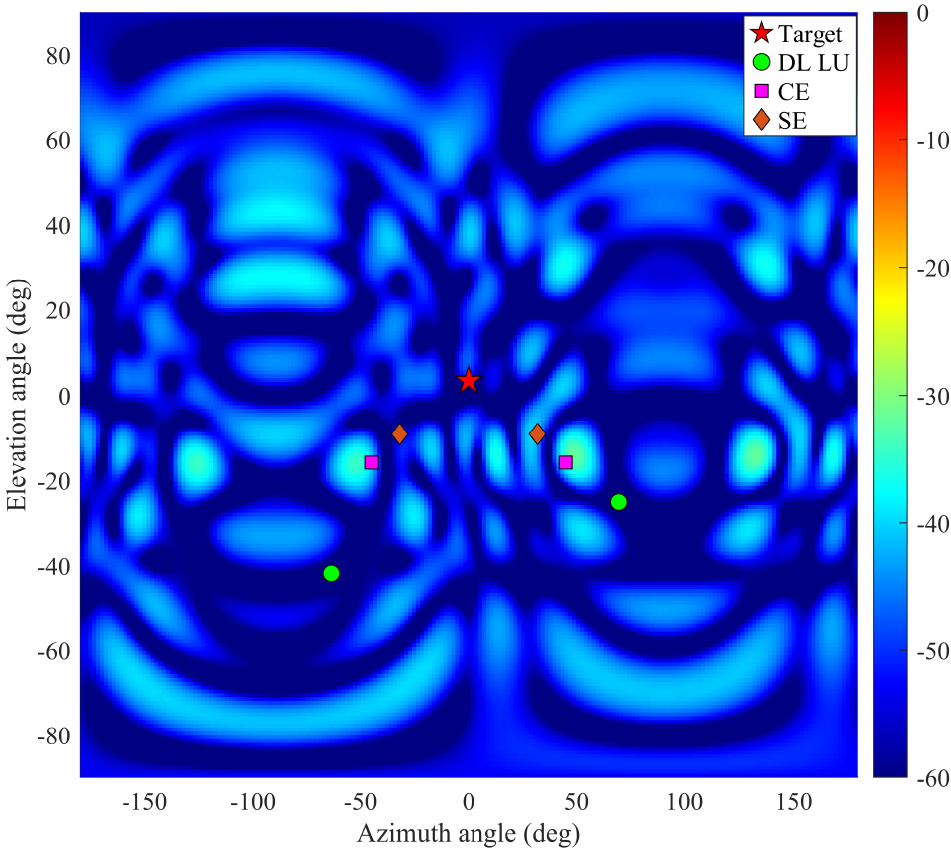}
    \label{AN_only}
  }
  \hfill  
  \subfloat[Aggregate signals]{
    \includegraphics[width=0.255\textwidth]{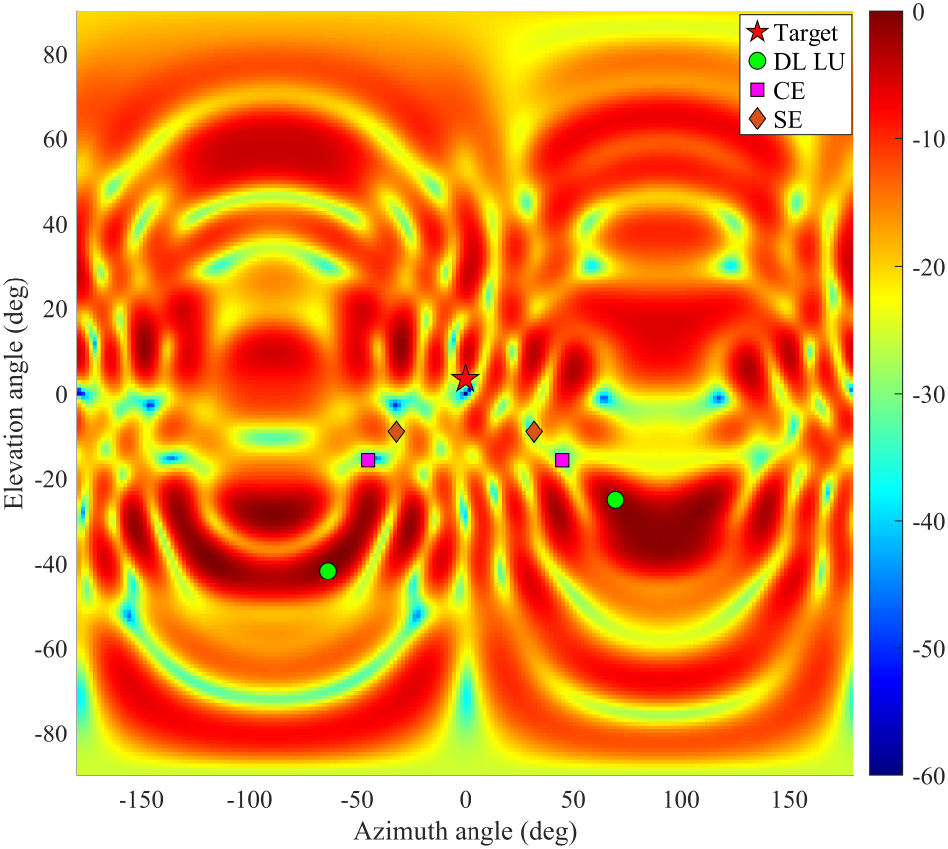}
    \label{transmit_all}
  }
  \vspace{-0.35em}
  \caption{Transmit beampattern ($N=12$, unit: dB).}
  \label{transmit_beampattern}
  \vspace{-2em}
\end{figure*}

\begin{figure*}[t]
	\centering
	\subfloat[Receive beampattern for the target]{
		\includegraphics[width=0.255\textwidth]{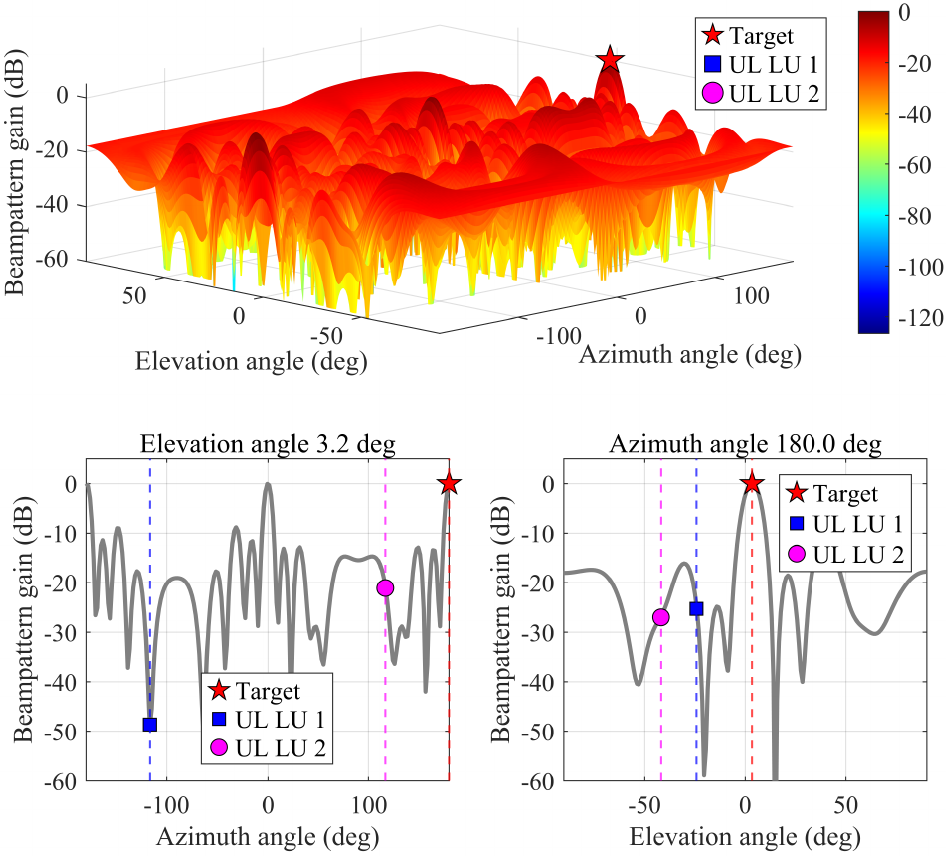}
		\label{receive_target}
	}
	\hfill  
	\subfloat[Receive beampattern for UL LU 1]{
		\includegraphics[width=0.255\textwidth]{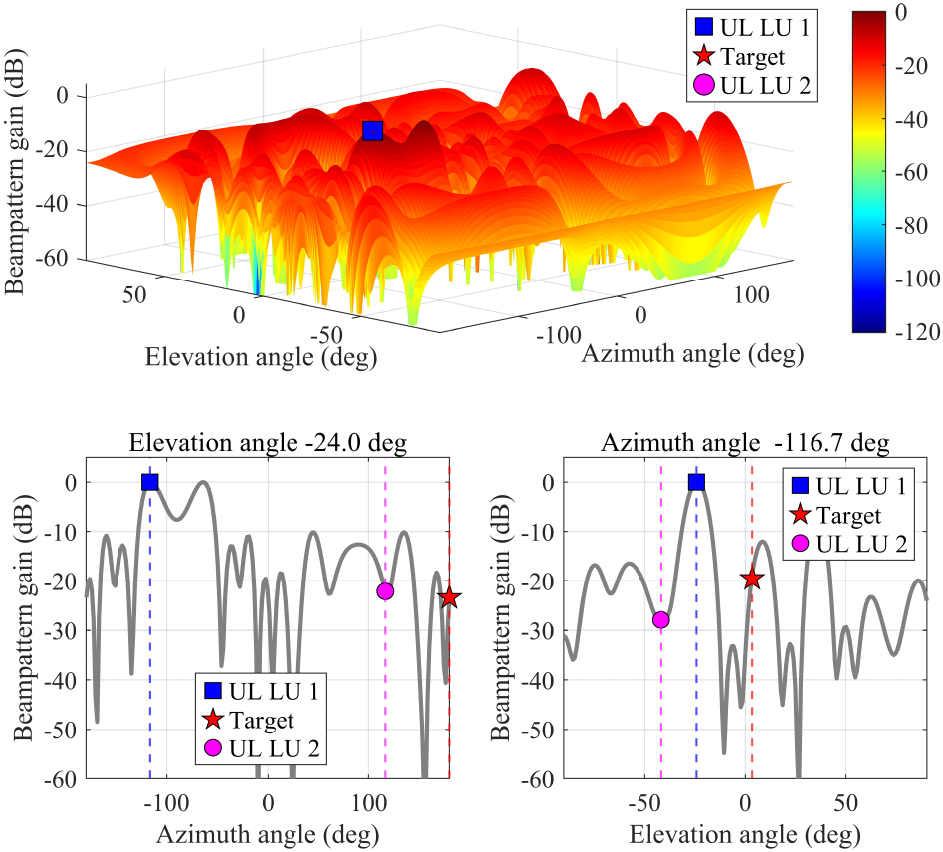}
		\label{receive_UE1}
	}
	\hfill  
	\subfloat[Receive beampattern for UL LU 2]{
		\includegraphics[width=0.255\textwidth]{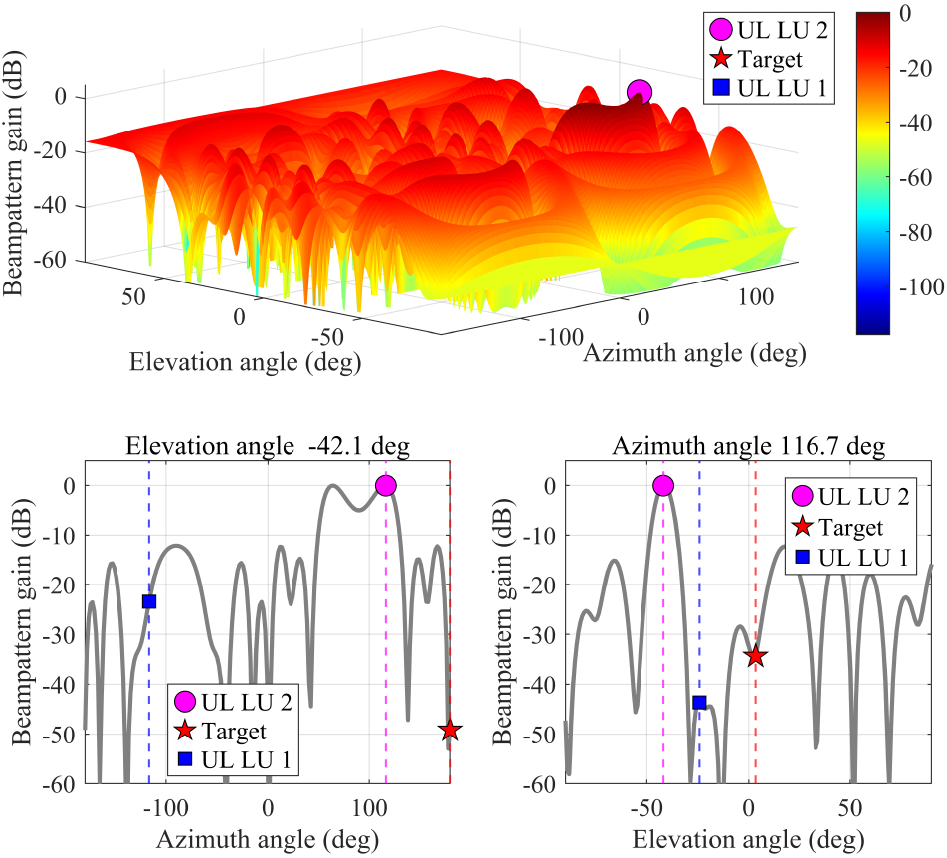}
		\label{receive_UE2}
	}
    \vspace{-0.35em}
	\caption{Receive beampattern ($N=12$, unit: dB).}
	\label{receive_beampattern}
	\vspace{-2em}
\end{figure*}

\begin{figure*}[t]
	\centering
	\subfloat[]{
		\includegraphics[width=0.255\textwidth]{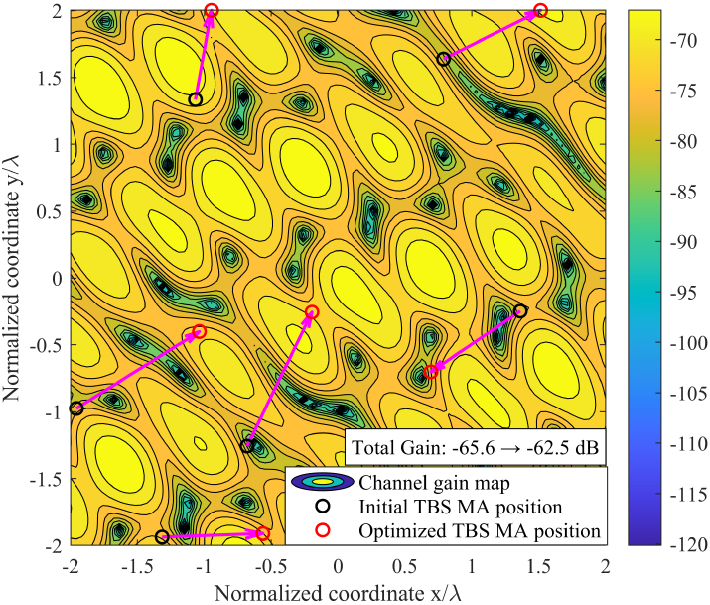}
		\label{DL_channel_map}
	}
	\hfill  
	\subfloat[]{
		\includegraphics[width=0.255\textwidth]{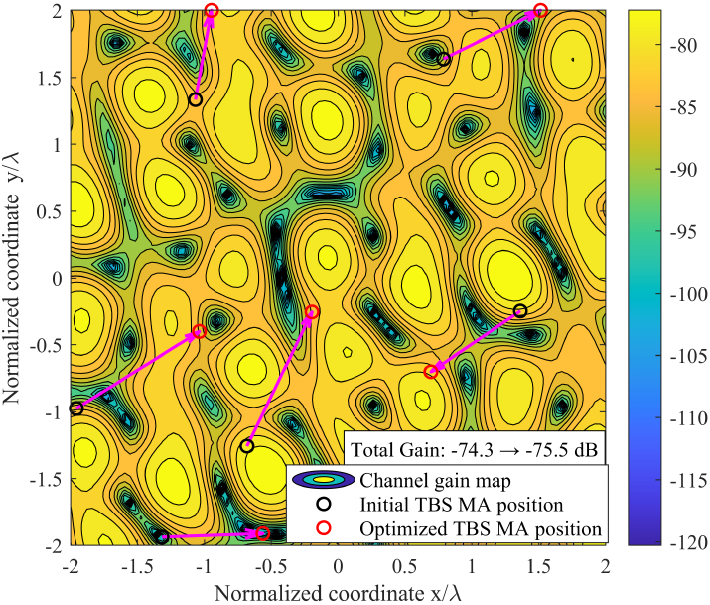}
		\label{EVE_channel_map}
	}
	\hfill  
	\subfloat[]{
		\includegraphics[width=0.255\textwidth]{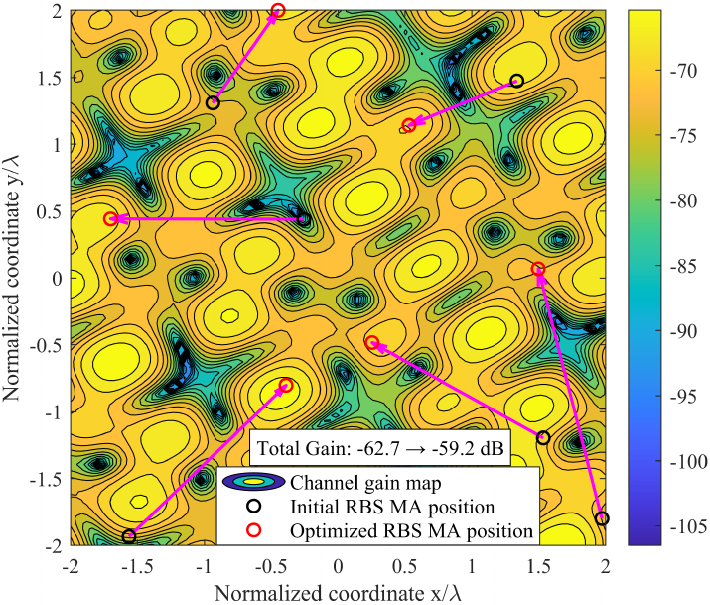}
		\label{UL_channel_map}
	}
    \vspace{-0.35em}
	\caption{Channel power gains (unit: dB) (a) between transmit MA of the TBS and single FPA of the DL LU, (b) between transmit MA of the TBS and single FPA of the CE, (c) between single FPA of the UL LU and receive MA of the RBS.}
	\label{channel_map}
	\vspace{-1.5em}
\end{figure*}

To visualize the beamforming capabilities, Fig.~\ref{transmit_beampattern} and Fig.~\ref{receive_beampattern} present the realized beampattern gains (in dB). We consider a simplified topology with $N=12$ antennas per BS. The TBS and RBS are deployed at $(0, 50, 20)$ m and $(100, 50, 20)$ m, respectively. The target is located at $(50, 50, 23)$ m, surrounded by clutter sources. The DL LUs are positioned at $(10, 30, 0)$ m and $(15, 90, 0)$ m, while UL LUs are located at $(80, 10, 0)$ m and $(90, 70, 0)$ m. CEs are placed at $(50, 0, 0)$ m and $(50, 100, 0)$ m, with SEs at $(80, 0, 5)$ m and $(80, 100, 5)$ m.

Fig.~\ref{transmit_beampattern} illustrates the normalized transmit beampatterns. As shown in Fig.~\ref{transmit_beampattern}(a), the communication-only beampattern directs its high gains toward the DL LUs while exhibiting substantially lower responses toward the CEs and SEs, thereby enhancing desired signal delivery and limiting potential information leakage. In contrast, the AN-only beampattern in Fig.~\ref{transmit_beampattern}(b) exhibits stronger responses toward the eavesdropper directions while maintaining low radiation levels toward the DL LUs. This enables AN to impair eavesdropping reception while mitigating its interference with legitimate communications. The aggregate beampattern in Fig.~\ref{transmit_beampattern}(c) combines these effects, maintaining strong radiation toward the DL LUs while providing sufficient illumination toward the target to satisfy the sensing MI requirement. These results demonstrate the effectiveness of the proposed joint beamforming design in enhancing communication secrecy through directional transmission and AN-assisted interference, while satisfying the sensing performance and sensing eavesdropping constraints.

Subsequently, we investigate the normalized receive beamforming performance at the RBS in Fig.~\ref{receive_beampattern}. As shown in Fig.~\ref{receive_beampattern}(a), the sensing receive beampattern exhibits a high-gain response toward the target while significantly attenuating the responses toward UL LUs. This facilitates target echo reception while mitigating interference from UL transmissions. Similarly, Figs.~\ref{receive_beampattern}(b) and~\ref{receive_beampattern}(c) show that the receive beamformers for UL LU 1 and UL LU 2 provide strong responses toward their respective desired directions while suppressing target echoes and interference from the other UL LU. This enhances the reception of desired UL signals and improves the legitimate UL SINRs, thereby contributing to improved UL secrecy performance. Together with the transmit beampatterns, these results demonstrate the effectiveness of the proposed joint beamforming design in supporting secure communications and reliable target sensing.

To further elucidate how MA position optimization exploits spatial channel variations, Fig.~\ref{channel_map} visualizes the channel power gain realizations relative to the spatial distribution of MAs. For clarity, we focus on a representative scenario involving a DL LU, a UL LU, and a CE. As observed in Figs. \ref{channel_map}(a) and \ref{channel_map}(c), the optimized MA positions favor regions with higher channel gains, yielding substantial gain improvements of $3.1$ dB (from $-65.6$ dB to $-62.5$ dB) for the TBS-to-DL LU link and $3.5$ dB (from $-62.7$ dB to $-59.2$ dB) for the UL LU-to-RBS link. Conversely, Fig.~\ref{channel_map}(b) shows that the optimized MA positions reduce the TBS-to-CE channel gain by $1.2$ dB (from $-74.3$ dB to $-75.5$ dB), thereby weakening the eavesdropping channel. These results demonstrate that the proposed scheme effectively leverages the spatial DoFs of MAs to enhance legitimate channel gains while suppressing the eavesdropping channel gain, creating more favorable channel conditions for secure communications. Furthermore, the distinct spatial separation of optimized MA positions confirms adherence to the minimum inter-MA spacing constraints.

In the simulations, our proposed scheme is denoted as ``Proposed, MA''. To fully evaluate its advantages, we compare it against the following benchmarks: (1) {Standard PSO, MA}: Optimizes MA positions via standard PSO. (2) {ZF, MA}: Employs Zero-Forcing (ZF) receivers at the RBSs, replacing the optimal combiners derived in \eqref{sensing_filter} and \eqref{UL_filter}. (3) {NoAN, MA}: Transmits only confidential signals without AN, allocating the full power budget for information beamforming. (4) {Random Position Antenna (RPA)} \cite{10243545}: Selects the configuration achieving the highest SSR from 100 random feasible realizations. (5) {FPA}: Replaces MAs with fixed UPAs centered within the designated moving regions.

\begin{figure}[t]
	\centering
		\includegraphics[height=45.7mm]{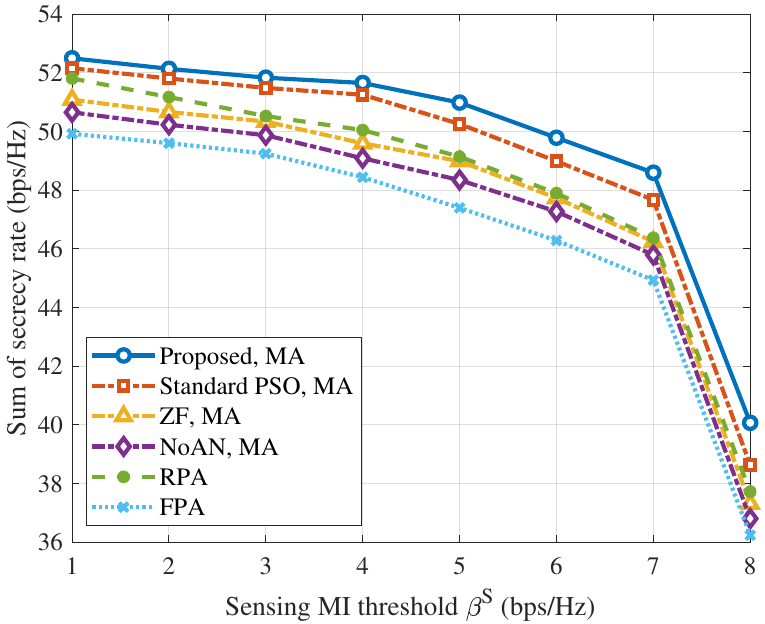}
		\vspace{-0.35em}
		\caption{SSR versus sensing MI.}
        \vspace{-1em}
		\label{SSR_MI}
\end{figure}

Fig.~\ref{SSR_MI} illustrates the impact of sensing MI threshold $\beta^\mathrm{S}$ on the SSR. The SSR decreases monotonically across all schemes as $\beta^\mathrm{S}$ increases, highlighting the inherent trade-off between sensing performance and communication secrecy. Higher sensing MI requirements impose more stringent constraints on joint beamforming, power allocation, and MA positioning, thereby limiting the flexibility available for secrecy rate optimization. Specifically, as sensing constraints become stringent (e.g., $\beta^\mathrm{S} > 7$ bps/Hz), the SSR drops precipitously due to exacerbated resource contention where prioritizing sensing compromises security. Nevertheless, the proposed algorithm consistently outperforms all benchmarks across the considered range of sensing MI thresholds. These results demonstrate the effectiveness of the proposed joint optimization framework in maintaining superior secrecy performance under increasingly stringent sensing requirements.

\begin{figure}[t]
	\centering
		\includegraphics[height=45.7mm]{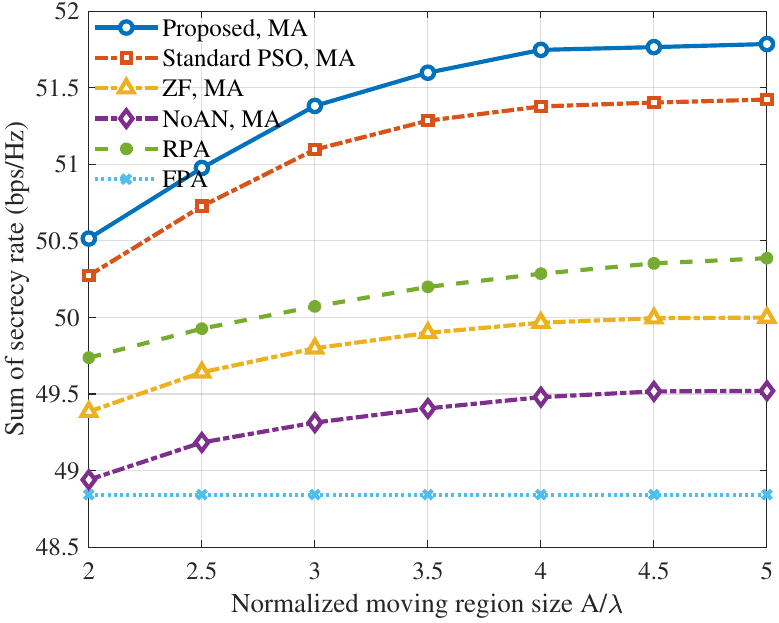}
		\vspace{-0.35em}
		\caption{SSR versus normalized moving region size.}
		\vspace{-1em}
		\label{fig:SSR_size}
\end{figure}

Fig.~\ref{fig:SSR_size} presents the SSR versus the normalized moving region size $A/\lambda$. It can be observed that the proposed algorithm consistently outperforms the benchmarks, with the SSRs of all MA-assisted schemes increasing monotonically before gradually leveling off. This trend is attributed to the fact that larger regions provide greater flexibility for MA positioning, enabling MAs to more effectively exploit spatial channel variations. However, saturation occurs as further increases in the moving region size offer diminishing opportunities for channel gain improvement. In contrast, the SSR of the FPA scheme remains unchanged because its antenna positions are fixed and thus unaffected by the moving region size. Notably, the proposed SA-MCPSO outperforms standard PSO, particularly as the search space expands, by leveraging multiple candidate velocity updates to improve the exploration of MA positions. Furthermore, while outperforming FPA, the RPA scheme yields limited gains because it selects MA positions without leveraging CSI. Additionally, the lower SSR achieved by the NoAN scheme highlights the importance of AN in impairing eavesdropper reception and enhancing communication secrecy.

\begin{figure}[t]
	\centering
		\includegraphics[height=45.7mm]{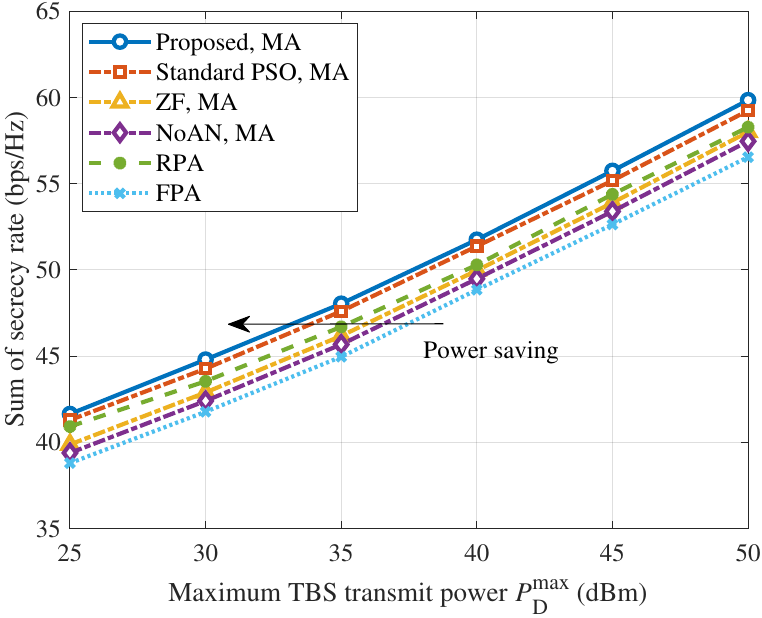}
		\vspace{-0.35em}
		\caption{SSR versus TBS transmission power.}
		\vspace{-1em}
		\label{fig:SSR_P}
\end{figure}

The impact of the maximum TBS transmit power $P_{\mathrm{D}}^{\max}$ is examined in Fig.~\ref{fig:SSR_P}. The SSR of all schemes exhibits monotonic growth, as a higher transmit power budget provides greater flexibility for enhancing legitimate signal reception and managing interference. The proposed algorithm consistently outperforms the benchmarks across the considered power range, maintaining a performance advantage over Standard PSO, RPA, and FPA. Notably, as indicated by the arrow, the proposed scheme achieves the same SSR under a lower transmit power budget, indicating a potential power-saving advantage. These results validate the effectiveness of joint MA positioning and beamforming in enhancing communication secrecy under limited transmit power budgets.

\begin{figure}[t]
	\centering
		\includegraphics[height=45.7mm]{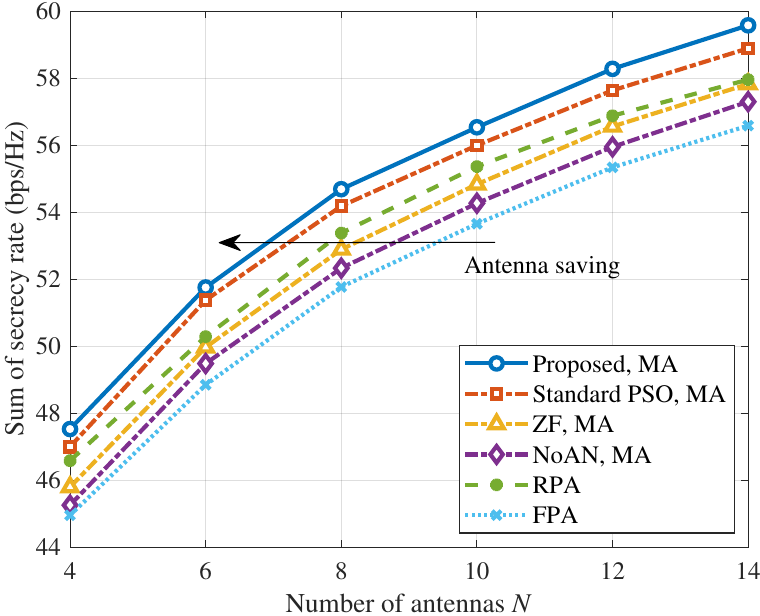}
		\vspace{-0.35em}
		\caption{SSR versus the number of antennas.}
		\vspace{-1em}
		\label{fig:SSR_N}
\end{figure}

Fig.~\ref{fig:SSR_N} reveals the performance improvements yielded by increasing the number of antennas $N$. A monotonic increase in SSR is observed for all considered schemes, driven by the enhanced beamforming gain and spatial diversity provided by larger arrays, which facilitate desired signal enhancement and interference suppression. MA-based schemes consistently outperform the FPA benchmark, with the performance gap generally widening as $N$ increases. This advantage arises because FPA is restricted to fixed antenna positions, whereas MAs provide greater flexibility to exploit favorable spatial channel conditions. Among the MA-based approaches, the proposed algorithm achieves the highest SSR. Furthermore, as highlighted by the arrow, the proposed scheme achieves a given SSR with fewer antennas than the benchmarks. This suggests that optimizing MA positions can improve the utilization of antenna resources and reduce the dependence on larger antenna arrays, thereby underscoring its practical superiority.

\begin{figure}[t]
	\centering
		\includegraphics[height=45.7mm]{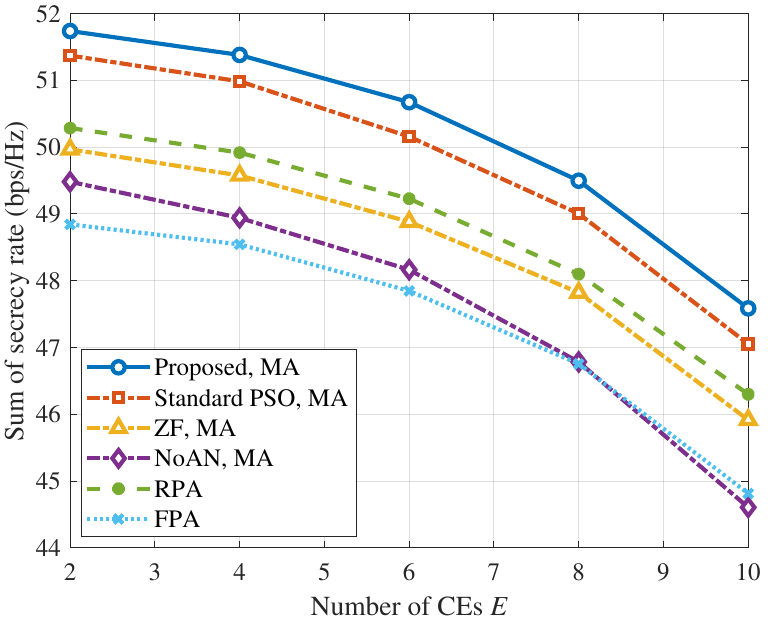}
		\vspace{-0.35em}
		\caption{SSR versus the number of CEs.}
		\vspace{-1em}
		\label{fig:SSR_CE}
\end{figure}

\begin{figure}[t]
	\centering
		\includegraphics[height=45.7mm]{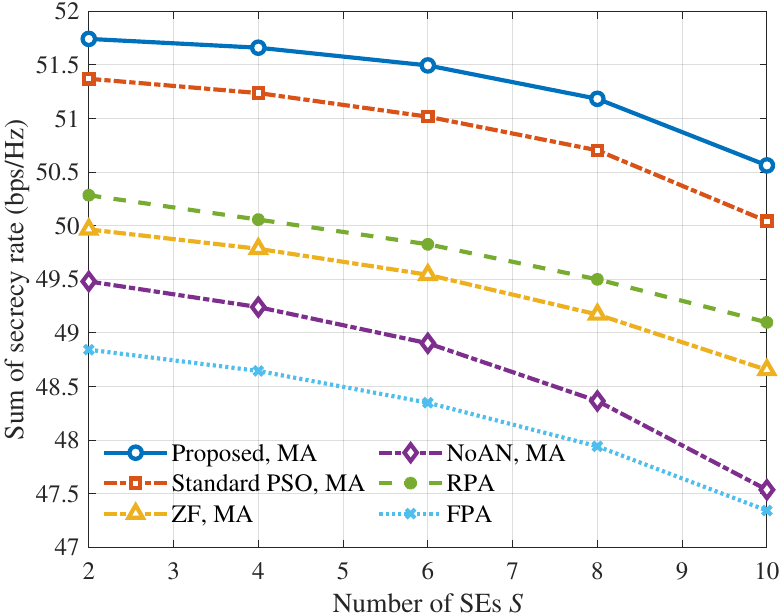}
		\vspace{-0.35em}
		\caption{SSR versus the number of SEs.}
		\vspace{-1em}
		\label{fig:SSR_SE}
\end{figure}

Fig.~\ref{fig:SSR_CE} investigates the impact of the number of CEs $E$ on the SSR. As $E$ increases, the SSR decreases monotonically for all schemes. This is because the CEs are assumed to cooperatively intercept the confidential UL and DL signals, and thus a larger $E$ strengthens the aggregate eavesdropping capability and directly reduces the achievable SSR. Despite the increasingly severe cooperative eavesdropping, the proposed scheme consistently achieves the highest SSR over the entire range of $E$. In particular, its persistent advantage over Standard PSO demonstrates the effectiveness of the proposed SA-MCPSO in identifying more favorable antenna configurations as the number of CEs increases. The performance loss of the NoAN scheme becomes more evident as $E$ grows, since the absence of AN weakens the system's capability to degrade the received signal quality at the CEs. Moreover, RPA provides only limited improvement over FPA because it searches over randomly generated feasible antenna configurations rather than adaptively optimizing the MA positions to exploit channel-dependent spatial variations. The ZF-based scheme also achieves a lower SSR than the proposed design because ZF prioritizes interference nulling and may sacrifice desired-signal gain while amplifying noise, whereas the proposed receive combiners better balance desired-signal preservation and interference-plus-noise suppression. These results confirm that the proposed joint design effectively mitigates the SSR degradation caused by an increasing number of CEs.

Fig.~\ref{fig:SSR_SE} shows the SSR as the number of SEs $S$ increases. A monotonic SSR reduction is again observed for all schemes. The underlying reason is that each SE imposes an individual constraint on its target-detection probability, and the accumulation of these constraints progressively contracts the feasible region of the joint design. Therefore, the transmit covariance and MA positions must be coordinated more carefully to simultaneously satisfy the prescribed sensing MI and the target-detection probability constraints of multiple SEs. The proposed scheme maintains a clear performance advantage across the considered range of $S$. The inferior performance of NoAN indicates that removing AN reduces the flexibility of the transmit design in simultaneously suppressing communication eavesdropping and controlling the energy observed by the SEs. RPA and FPA are limited by random or fixed antenna configurations and therefore cannot fully capitalize on channel-dependent spatial variations. The ZF-based scheme also yields a lower SSR because its strict interference-nulling strategy sacrifices the tradeoff between desired-signal enhancement and interference-plus-noise suppression. Compared with Fig.~\ref{fig:SSR_CE}, the SSR degradation caused by increasing $S$ is less pronounced than that caused by increasing $E$, since additional CEs directly strengthen the eavesdropping terms in the SSR, whereas additional SEs affect the SSR through additional sensing-security constraints. This comparison further demonstrates the effectiveness of the proposed scheme in mitigating the adverse impact of increasing numbers of CEs and SEs on the SSR.

\begin{figure}[t]
	\centering
		\includegraphics[height=45.7mm]{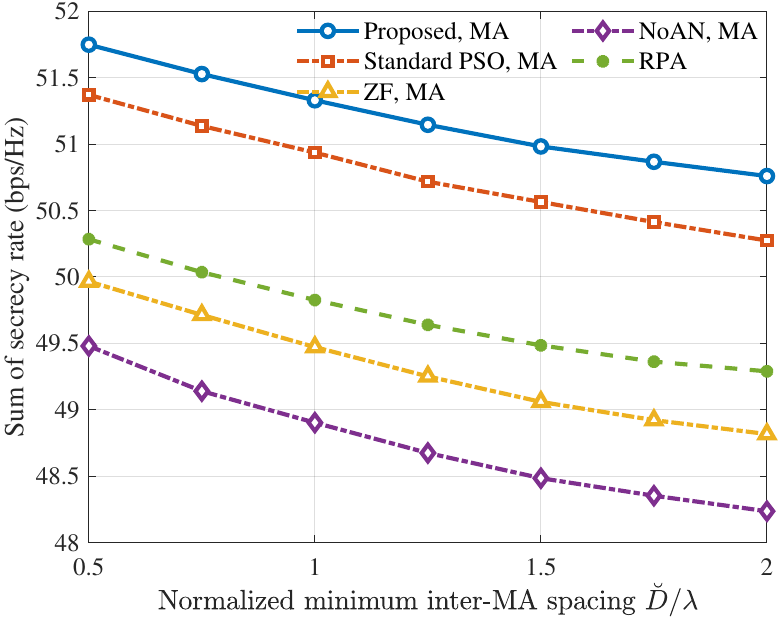}
		\vspace{-0.35em}
		\caption{SSR versus the normalized minimum inter-MA spacing.}
		\vspace{-1em}
		\label{fig:SSR_space_in_MA}
\end{figure}

Fig.~\ref{fig:SSR_space_in_MA} illustrates the impact of the normalized minimum inter-MA spacing $\breve{D}/\lambda$ on the SSR. The FPA benchmark is omitted because its antenna locations and inter-element spacing are fixed. As $\breve{D}/\lambda$ increases from $0.5$ to $2$, the SSR decreases for all MA-enabled schemes. A larger minimum spacing imposes a more restrictive geometric constraint on the MA configuration, thereby shrinking the feasible position set and limiting the exploitation of continuous spatial DoFs. As a result, the MAs have less flexibility to independently adjust their positions to exploit favorable channel conditions for strengthening legitimate links and suppressing eavesdropping, which leads to a reduction in the achievable SSR. The proposed scheme consistently attains the highest SSR over the entire spacing range, demonstrating its effectiveness in exploiting the remaining spatial DoFs under increasingly restrictive MA-spacing constraints. Its advantage over Standard PSO further indicates that the proposed SA-MCPSO can more effectively search the constrained MA configuration space. RPA achieves only limited gains because it relies on randomly sampled feasible antenna configurations rather than adaptively optimizing the MA positions according to the channel conditions. The lower SSRs of ZF and NoAN further reflect the performance loss associated with strict interference nulling and the reduced eavesdropper-suppression capability without AN, respectively. These results demonstrate that increasing the minimum inter-MA spacing restricts the spatial reconfiguration capability of MAs, while the proposed joint design can partially alleviate the resulting SSR degradation over the considered spacing range.

\begin{figure}[t!]
	\centering
		\includegraphics[height=45.7mm]{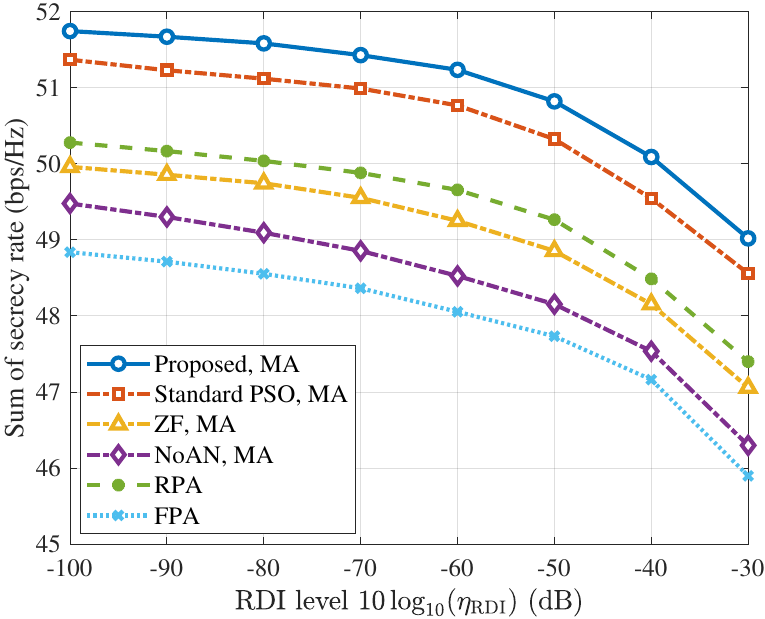}
		\vspace{-0.35em}
		\caption{SSR versus RDI level.}
		\vspace{-1em}
		\label{fig:SSR_HSI}
\end{figure}

To evaluate the impact of imperfect direct-link interference cancellation, we further consider the RDI from the TBSs to the RBSs. Let $\mathbf{H}^{\mathrm{TR}}$ denote the aggregate TBS-to-RBS direct-link channel. The RDI covariance is modeled as $\mathbf{C}_{\mathrm{RDI}}=\eta_{\mathrm{RDI}}\mathbf{H}^{\mathrm{TR}}\mathbf{R}(\mathbf{H}^{\mathrm{TR}})^H$, where $\eta_{\mathrm{RDI}}\in[0,1]$ denotes the fraction of direct-link interference power remaining after cancellation, with a smaller $\eta_{\mathrm{RDI}}$ corresponding to more effective interference cancellation. The resulting $\mathbf{C}_{\mathrm{RDI}}$ is incorporated into the interference covariance at the RBSs for both UL reception and target sensing.
Fig.~\ref{fig:SSR_HSI} illustrates the impact of RDI on the SSR, where the RDI level is characterized by $10\log_{10}(\eta_{\mathrm{RDI}})$. As the RDI level increases, the SSR decreases for all considered schemes. This is because stronger RDI increases the interference power at the RBSs, thereby degrading both UL reception and target sensing. Consequently, the sensing MI constraint becomes more difficult to satisfy, while the degraded UL reception directly reduces the achievable SSR.
The proposed scheme consistently achieves the highest SSR over the entire RDI range. The performance degradation becomes more pronounced at higher RDI levels, highlighting the importance of effective direct-link interference cancellation in networked ISAC systems. Nevertheless, the persistent advantage of the proposed scheme shows that the proposed joint design can partially mitigate the adverse impact of RDI by exploiting the available spatial DoFs. These results demonstrate that the proposed scheme retains a clear SSR advantage under non-ideal TBS-to-RBS direct-link interference cancellation.

\begin{figure*}[t]
	\centering
	\subfloat[]{
		\includegraphics[width=0.285\textwidth]{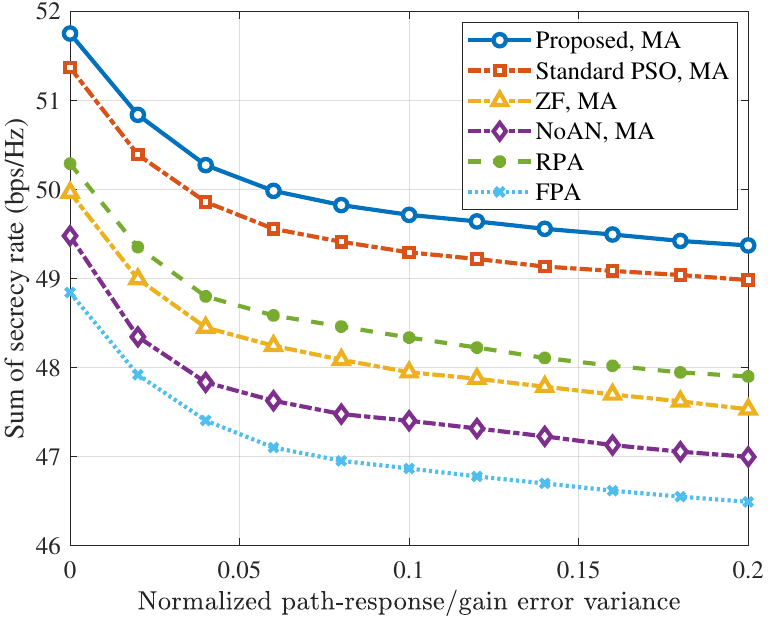}
		\label{SSR_PRV_error}
	}
	\hfill  
	\subfloat[]{
		\includegraphics[width=0.285\textwidth]{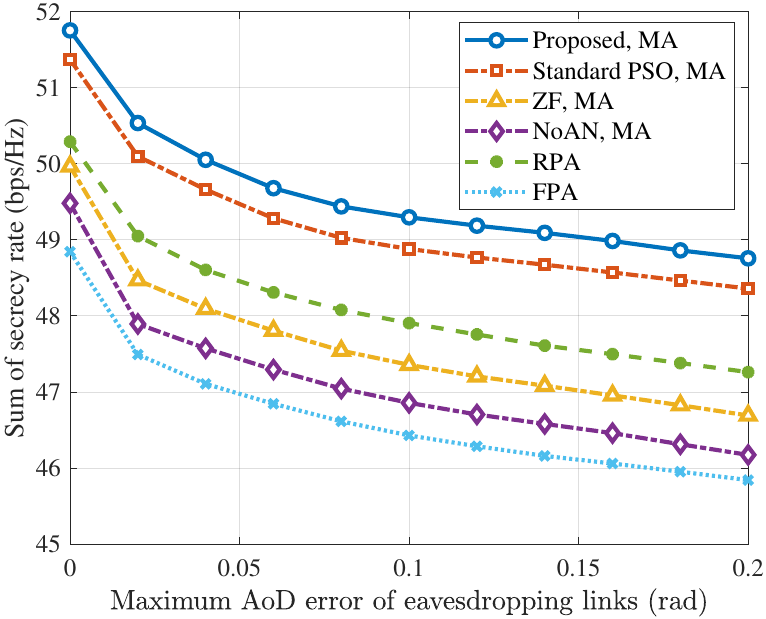}
		\label{SSR_AoD_error}
	}
	\hfill  
	\subfloat[]{
		\includegraphics[width=0.285\textwidth]{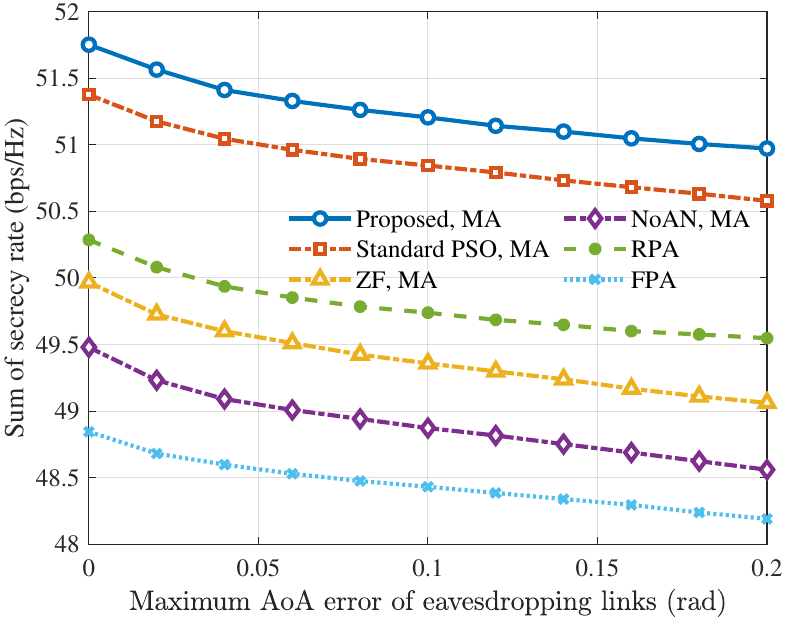}
		\label{SSR_AoA_error}
	}
   \vspace{-0.35em}
	\caption{Impact of imperfect eavesdropper CSI on the SSR: (a) normalized path-response/gain error variance, (b) maximum AoD error, and (c) maximum AoA error.}
	\label{SSR_CSI_error}
	\vspace{-1.55em}
\end{figure*}

Fig.~\ref{SSR_CSI_error} evaluates the sensitivity of the SSR to imperfect eavesdropper CSI. The channel information associated with the LUs and the target is assumed to be perfectly known, whereas the CSI of both CEs and SEs is subject to estimation errors. Three representative CSI error types are considered: (a) normalized path-response/gain error variance, (b) maximum AoD error of the eavesdropping links, and (c) maximum AoA error of the eavesdropping links.
As shown in Fig.~\ref{SSR_CSI_error}(a)--(c), the SSR decreases for all schemes as the eavesdropper CSI uncertainty increases. This degradation arises because inaccurate eavesdropper CSI creates a mismatch between the estimated and actual eavesdropping channels, thereby reducing the effectiveness of transmit beamforming, AN design, and MA-position optimization in suppressing information interception at the CEs and limiting target detection at the SEs. Since the channel information associated with the LUs and the target remains accurately known, the legitimate communication and sensing links can still be effectively served, whereas the capability to exploit the spatial characteristics of the eavesdropping links gradually deteriorates. 
Among the three CSI error types, the AoD error causes the most pronounced SSR degradation, followed by the normalized path-response/gain error variance and the AoA error. Specifically, when the maximum AoD error reaches $0.2$ rad, the SSR of the proposed scheme decreases from approximately $51.75$ to $48.76$ bps/Hz, corresponding to a loss of about $5.8\%$. By comparison, a normalized path-response/gain error variance of $0.2$ reduces the SSR to approximately $49.37$ bps/Hz, corresponding to a loss of about $4.6\%$, whereas a maximum AoA error of $0.2$ rad results in a much smaller reduction to approximately $50.97$ bps/Hz. This difference can be attributed to the fact that AoD errors directly perturb the transmit-side spatial responses involved in beamforming, AN design, and transmit-MA positioning, thereby weakening the suppression of eavesdropping links. Path-response/gain errors distort the estimated path responses/gains and hence the reconstructed eavesdropping channels, which impairs beamforming, AN design, and MA position optimization. In contrast, AoA errors primarily perturb the SE-side receive spatial responses involved in the sensing eavesdropping model and therefore exert a comparatively weaker influence on the SSR.
Despite the CSI mismatch, the proposed scheme consistently achieves the highest SSR over the entire error range. The antenna positions of FPA and RPA are not continuously optimized according to the eavesdropper CSI, since their antenna configurations are fixed or randomly sampled, respectively. However, their achievable SSRs remain lower because they cannot fully exploit the MA-enabled position-domain spatial DoFs. The lower SSRs of NoAN and ZF further reflect the reduced eavesdropper suppression capability without AN and the performance loss associated with strict interference nulling, respectively. These results show that the proposed scheme exhibits graceful performance degradation while retaining a clear SSR advantage under imperfect eavesdropper CSI.

\vspace{-0.58em}
\section{Conclusion}\label{5}
In this paper, a novel MA-enabled networked cooperative ISAC framework has been proposed to simultaneously secure coexisting UL and DL communications against colluding CEs while protecting target sensing from multiple SEs. We have investigated the joint design of transmit and receive MA positions, receive combiners, transmit beamforming, AN covariance, and UL power allocation to maximize the SSR subject to sensing MI, SE detection probability, transmission power, and MA deployment constraints. To tackle the resulting highly non-convex problem, the original problem was decomposed into three tractable subproblems addressed iteratively within an AO framework. Within this framework, closed-form solutions were derived for the receive combiners, while SCA was employed to optimize transmit beamforming and power allocation, and a tailored SA-MCPSO algorithm was applied for MA positioning until convergence. The convergence of the proposed algorithm has been analyzed, and its computational complexity has been derived. Numerical results have validated the convergence and effectiveness of the proposed algorithms, demonstrating superior performance compared to other state-of-the-art benchmarks. Furthermore, simulations demonstrated that the optimized MA positions can enhance legitimate channel gains and reduce eavesdropping channel gains while satisfying sensing requirements. Finally, the proposed framework exhibits distinct superiority in achieving power-saving and antenna-saving gains, highlighting improved resource efficiency in secure networked ISAC systems without compromising dual-domain security and sensing performance.

\balance
\bibliographystyle{IEEEtran}


\end{document}